\documentclass[11pt]{article}

\usepackage{amssymb,amsmath,amsthm}	% Mathematische Symbole
\usepackage{a4wide}
\usepackage{graphicx}      										% Grafiken einbinden
\usepackage{tikz}          										% LaTeX-Grafiken erzeugen
\usepackage{url}
\usepackage{dsfont}
\usepackage{amsmath}
\usepackage{xspace}
\usepackage{mathtools}% lädt auch amsmath
\mathtoolsset{showonlyrefs}

\newtheorem{definition}{Definition}[]

\newtheorem{remark}{Remark}[]

\newtheorem{proposition}{Proposition}[]

\newcommand{\R}{\ensuremath{\mathbb{R}}}

\DeclareMathOperator*{\argmax}{argmax}
\DeclareMathOperator*{\argmin}{argmin}
\begin{document}
	
	%-- TITEL ----------------------------------------------------------%
	\title{A Macroscopic Portfolio Model:\\ From Rational Agents to Bounded Rationality}
	
	\author{Torsten Trimborn\footnote{IGPM, RWTH Aachen, Templergraben 55, 52056 Aachen, Germany, trimborn@igpm.rwth-aachen.de}}

	\maketitle

   %-- INHALTSVERZEICHNIS ----------------------------------------------------------%

\begin{abstract}
We introduce a microscopic model of interacting financial agents, where each agent is characterized by two portfolios; money invested in bonds and money invested in stocks. 
Furthermore, each agent is faced with an optimization problem in order to determine the optimal asset allocation. The stock price evolution is driven by the aggregated investment decision
of all agents. In fact, we are faced with a differential game since all agents aim to invest optimal. Mathematically such a problem is ill posed and we introduce the concept of Nash equilibrium solutions to ensure the existence of a solution. Especially, we denote an agent who solves this Nash equilibrium exactly a rational agent. As next step we use model predictive control to approximate the control problem. 
This enables us to derive a precise mathematical characterization of the degree of rationality of a financial agent. This is a novel concept in portfolio optimization and can be regarded as a general approach. 
In a second step we consider the case of a fully myopic agent, where we can solve the optimal investment decision of investors analytically. 
We select the running cost to be the expected missed revenue of an agent and we assume quadratic transaction costs. More precisely the expected revenues are determined by a combination of a fundamentalist or chartist strategy. Then we derive the mean field limit of the microscopic model in order to obtain a macroscopic portfolio model. The novelty in comparison to existent macroeconomic models in literature is that our model is derived from microeconomic dynamics. The resulting portfolio model is a three dimensional ODE system which enables us to derive analytical results. 
The conducted simulations reveal that the model shares many dynamical properties with existing models in literature. Thus, our model is able to replicate the most prominent features of financial markets,
namely booms and crashes. In the case of random fundamental prices the model is even able to reproduce fat tails in logarithmic stock price return data. 
Mathematically, the model can be regarded as the moment model of the recently introduced mesoscopic kinetic portfolio model \cite{trimborn2017portfolio}.\\

{\textbf{Keywords:} portfolio optimization, model predictive control, stock market, bounded rationality, crashes, booms} 
\end{abstract}

\clearpage
	
\section{Introduction }	

%intro

For many years the Efficient Market Hypothesis (EMH) by Eugene Fama \cite{fama1965behavior} has been the dominant paradigm for modeling asset pricing models.
Many famous theoretical models in finance such as Merton's optimal portfolio model \cite{merton1975optimum} or the Black and Scholes model \cite{jensen1972capital}
presume the correctness of the EMH. In the past years there has been a shift from rational and representative financial agents to bounded rational and heterogeneous agents \cite{hommes2002modeling}. 
The former notion of agents is in agreement with the EMH, whereas the latter one contradicts the EMH and has to be understood in the sense of Simon \cite{simon}. 
The drift away from the EMH has been supported by several empirical studies \cite{malkiel2003efficient, lehmann1990fads} and the financial crashes of the past decade \cite{kirman2011crisis, anand2010epidemics}.\\ \\ 
Bounded rational agents are widely used in econophysical asset pricing models, particularly in agent-based-computational financial market models.
These models usually consider a large number of interacting heterogeneous financial agents. These large complex systems are studied by means of Monte Carlo simulations.
Major contributions are for example the Levy-Levy-Solomon \cite{levy1994microscopic}, Cont-Bochaud \cite{cont2000herd} and the Lux-Marchesi \cite{lux1999scaling} model. The benefit of these models are first explanations for the existence of stylized facts, such as 
fat tails in asset returns or volatility clustering. They have even led to alternative market hypothesis, such as the adaptive market hypothesis by Lo \cite{lo2004adaptive}  or the interacting agent hypotheses by Lux \cite{lux1999scaling} as alternative to the EMH. 
 The disadvantage of these agent-based asset pricing models are the impracticability to apply analytical methods.
Furthermore, it has been shown in several studies that many agent-based models exhibit finite size effects \cite{trimborn2018sabcemm, egenter1999finite}.\\ \\
Another popular approach are simple low dimensional dynamic asset pricing models which usually consider two types of financial agents. In most cases one agent follows a chartist strategy and the other a fundamental strategy. With a chartist strategy we mean an investor who bases his decision on technical trading rules, whereas a fundamentalist originates his investments from deviations of fundamentals to the stock price. These models are often formulated as two dimensional difference equations or as ordinary differential equations (ODE). They all have in common that the stock price equation is driven by the excess demand of financial agents. In comparison to agent-based models, these models can be regarded as macroscopic, since they consider aggregated quantities. 
In literature there are numerous models of that kind, for example by Beja and Goldman \cite{beja1980dynamic}, Day and Huang \cite{day1990bulls}, Lux \cite{lux1995herd}, Brock and Hommes \cite{brock1997rational, brock1998heterogeneous}, Chiarella \cite{chiarella1992dynamics} and Franke and Westerhoff \cite{franke2012structural}.
These asset pricing models feature a rich body of complex phenomena, such as limit cycles, chaotic behavior and bifurcations. Economically, these models study for example the impact of behavioral and psychological factors such as risk tolerance on the price behavior. Furthermore, they discover the origins of stylized facts. More precisely they study the source of booms and crashes and excess volatility.\\ \\
The financial agents are always modeled as bounded rational agents in the sense of Simon \cite{simon} and possess behavioral factors.
The importance of psychological influences in agent modeling has been emphasized by several authors \cite{shiller2003efficient, kahneman2003maps, lux2008stochastic, conlisk1996bounded}. 
One may note that the precise form of agent demand is not established from microscopic dynamics and the connection to rational agents is unclear. 
Albeit the form of the agent demand in the models \cite{brock1998heterogeneous, chiarella2002heterogeneous} is derived by a mean-variance wealth maximization the expected stock return over bond return is modeled macroscopically. In addition, these models neglect the impact of wealth evolution on the agent demand. One example of a bounded rational agent is a myopic agent, who basis his action only on currently available informations \cite{brown1981myopic}. To our knowledge, a precise mathematical notion of rationality in the context of portfolio optimization is missing. In addition, there is a lack of explanation concerning the interrelations between the action of a rational agent and for example a fully myopic agent. \\ \\
For these reasons we introduce a rather general mathematical framework in order to quantify the level of rationality of each agent in the context of portfolio optimization. 
More precisely, we formulate a model of rational agents on the microscopic level. Thus, each agent is faced with an optimal control problem in order to optimize their portfolio and determine the optimal investment decision. In fact, each agents' portfolio dynamics is divided in the time evolution of the two asset classes bonds and stocks. We employ the notion of 
Nash equilibrium solutions \cite{bressan2011noncooperative} to ensure that the optimization problem is well posed. In addition, we define a rational agent to be an agent who solves the differential game exactly, respectively obtains the Nash equilibrium solution. In a next step we apply model predictive control \cite{camacho2013model} to approximate the optimization problem. This enables us to give a precise mathematical definition of the level of rationality of an investor. Especially, thanks to this methodology we obtain a natural connections between rational and bounded rational agents. Up to this step the approach is fully generic since each agent is equipped with 
rather general wealth dynamics. The stock price equation is driven by the aggregated excess demand of agents in agreement with the Beja-Goldman \cite{beja1980dynamic} or Day-Huang \cite{day1990bulls} model. In a second step we consider the situation of a fully myopic agent which enables us to compute the optimal control explicitly. We model the cost function to measure the expected lost revenue of the investor. The return estimate is a convex combination of a pure chartist or pure fundamental trading strategy. 
The weight between both strategies is determined by an instantaneous comparison of the chartist and fundamental return estimate and is closely connected to the strategy change in the Lux-Marchesi model \cite{lux1999scaling}. We obtain a large dynamical system in the spirit of known agent-based financial market models. The disadvantage is that such models are far to complex to study them by analytical methods. For that reason we derive the mean field limit \cite{stanley1971phase, golse2016dynamics} of our system. More precisely we derive the time evolution of the average money invested in stocks and the average money invested in bonds. Hence, we obtain a macroeconomic portfolio model of three ODEs. This time continuous model shares many similarities with macroeconomic models in literature e.g. with the model by Lux  \cite{lux1995herd}. The novelty of our approach is that the resulting macroeconomic model is supported by precise microeconomic dynamics.
In general the advantage of a time continuous model is the possibility to use many analytical tools in order to quantify the dynamic behavior of the model. A second advantage is that the model can be studied on arbitrary time scales since we consider dimensionless quantities.  \\ \\
%summary of outline of paper
The outline of this paper is as follows: First we consider microscopic agent dynamics. In fact, we first introduce a microscopic model of rational agents. As second step, we approximate the complex optimization problem and give a connection between rational and bounded rational agents. Then finally, we derive the macroscopic portfolio model in the case of fully myopic agents. In section 4 we study the qualitative behavior of our model. Thus, we discuss possible steady states and study the dynamics caused by a pure fundamental or pure chartist strategy. In the following section, we present numerous simulation results and analyze the impact of several model parameters. Finally, we give a conclusion and short discussion of our results.   
\section{Economic Microfoundations}
We consider $N$ financial agents equipped with their personal monetary wealth $w_i\geq 0$. We denote all microscopic quantities with small letters. The non-negativity condition means that no debts are allowed. 
This wealth is divided in the wealth invested in the asset class stocks $x_i(t)>0$ and the wealth invested in bonds $y_i(t)>0$. We neglect all other asset classes and assume that $w_i(t)=x_i(t)+y_i(t)$ holds. 
The time evolution of the risk-free asset is described by a fixed non-negative interest rate $r\geq 0$ and the evolution of the risky asset by the stock return,
\[
\frac{\dot S(t)+D(t)}{S(t)},
\]
where $S(t)$ is the stock price at time $t$ and $D(t)\geq 0$ the dividend. 
The agent can shift capital between the two assets. We denote the shift from bonds into stocks by $u_i$. Thus, we have the dynamics
\begin{align*}
& \dot{x}_i(t) = \frac{\dot S(t)+D(t)}{S(t)}\ x_i(t)+u_i(t)\\
& \dot{y}_i(t) = r\ y_i(t)-u_i(t).
\end{align*}
Notice, that $u_i$ determines the investment decision of agents and implicitly specifies the asset allocation between both portfolios. 
We still need to describe the time evolution of the stock price $S$. 
We define the aggregated excess demand $ED_N$ of all financial agents as the average of all investment decisions of the agents.
$$
ED_N(t) := \frac{1}{N} \sum\limits_{i=1}^N u_i(t).
$$ 
The aggregated demand $ED_N$ is the average of agents' excess demand $u_i$. The agents' excess demand $u_i$ was defined as the investment decision of agents and can be interpreted as the  demand minus the supply of each agent. Hence, the excess demand is positive if the investors buy more stocks than they sell. For further details regarding the aggregated excess demand we refer to \cite{mantel1974characterization, sonnenschein1972market, cont2000herd, zhou2007self, trimborn2018sabcemm}.
Thus, the macroscopic \textbf{stock price evolution} is driven by the excess demand and is given by
\begin{align}
\dot{S}(t)= \kappa\ ED_N(t)\ S(t).\label{MacroStockODE}
\end{align}
where the constant $\kappa>0$ measures the market depth. This model for the stock price is commonly accepted \cite{beja1980dynamic, day1990bulls, zhou2007self, lux1995herd, hommes2006heterogeneous, trimborn2018sabcemm}. The ODE \eqref{MacroStockODE} can be interpreted as a relaxation of the well known equilibrium law, supply equals demand, dating back to the economist Walras \cite{walras1898etudes}.\\

\paragraph{Microscopic portfolio optimization}
As in classical economic theory, $u_i$ will be a solution of a risk or cost minimization.
The precise model of the objective function $\Psi_i(t)$ is left open at this point. We want to emphasize that $\Psi_i$ may depend on the stock price $S$ and the wealth of the agent's portfolios. 
The agent tries to minimize the running costs 
$$
\int_0^T \left( \frac{\mu}{2}u_i(t)^2+\Psi_i(t)\right) dt.
$$
We consider a finite time interval $[0,T]$ and have added a penalty term that punishes transactions. The penalty term is necessary to convexify the problem but is also reasonable, because it describes transaction costs. The transaction costs are modeled to be quadratic which is a frequently used assumption in portfolio optimization \cite{bertsimas2008robust, mitchell2013rebalancing, niehans1992international, gros1987effectiveness}. Furthermore, we assume that there are no final costs at final time $T$ present. \\
Hence, in summary, the microscopic model is given by

\begin{align}
& \dot{x}_i(t) = \frac{\dot S(t)+D(t)}{S(t)}\ x_i(t)+u^*_i(t)\\
& \dot{y}_i(t) = r\ y_i(t)-u^*_i(t)\\\label{microModelOpt}
& \dot{S}(t) = \kappa\ ED_N(t)\ S(t)\\ 
&u_i^*:= \argmax\limits_{u_i:[0,T]\to \R}  \int_0^T \left( \frac{\mu}{2}u_i(t)^2+\Psi_i(t)\right) dt. 
\end{align}
%\label{opt}
The microscopic model is an optimal control problem. The dynamics are strongly coupled by the stock price in a non-linear fashion.  In fact, it is impossible that all agents minimize their individual cost function since all agents play a game against each other. We are faced with a non-cooperative differential game. We choose the concept of Nash equilibria which will be explained in detail in the next section. 

\subsection{From Rational Agents to Bounded Rationality}
As we have defined the microscopic portfolio model we want to specify different solution of \eqref{microModelOpt} with respect to their economic interpretation.
We define the Nash equilibrium of a differential game. 
\begin{definition}
 A vector of control functions $t\mapsto (u_1^*,...,u_N^*)^T$ is a Nash equilibrium for a differential game
 
	\begin{align}
		\argmin\limits_{u_i:\ [0,T]\to \mathbb{R}} {q}_i(\boldsymbol{x}(T))-\int\limits_0^T {L_i}(t,\boldsymbol{x},\boldsymbol{u})\ dt, \label{MicroGame}
	\end{align}
	where $\boldsymbol{x}=(x_1,...,x_N)^T\in \R^{d N},\ x_i=(x_i^1,...,x_i^d)^T\in\R^d,\ 1\leq i\leq N,\ d\geq 1,\ \boldsymbol{u}=(u_1,...,u_N)^T\in \R^N$ and with the state dynamics
	\begin{align}
		\dot{\boldsymbol{x}}=\boldsymbol{f}(t,\boldsymbol{x},\boldsymbol{u}),\quad \boldsymbol{x}(0)=\boldsymbol{x}_0\in\R^{dN}, \label{stateDyn}
	\end{align}
with $\boldsymbol{f}=({f}_1,...,{f}_N)^T,\ \boldsymbol{L}=({L}_1,...,{L}_N)^T,\ \boldsymbol{q}=({q}_1,...,{q}_N)^T\ $ are defined on:
\begin{align*}
&{f}_i: [0,T]\times \R^{dN}\times \R^N\to \R^d,\\
&{L}_i: [0,T]\times \R^{dN}\times \R^N\to \R,\\
&{q}_i:\ \R^{dN}\to\R.
\end{align*}	 
 in the class of open-loop strategies if the following holds.\\
The control $u_i^*$ provides a solution to the optimal control problem for player i:
$$
		\argmin\limits_{u_i: [0,T]\to\R} q_i(\boldsymbol{x}(T))-\int\limits_0^T L_i(t,\boldsymbol{x},u_i, \boldsymbol{u}^*_{-i})\ dt,
$$
with the dynamics \eqref{stateDyn}. Here, $L_i$ denotes the running cost and $q_i$ the terminal cost. 

\end{definition}
For details regarding differential games and the idea of Nash equilibrium solutions we refer to \cite{bressan2011noncooperative}. 
This mathematical equilibrium concept enables us to give a precise definition of rationality in economics.

\begin{figure}[h!]
\begin{center}
\includegraphics[width=0.8\textwidth]{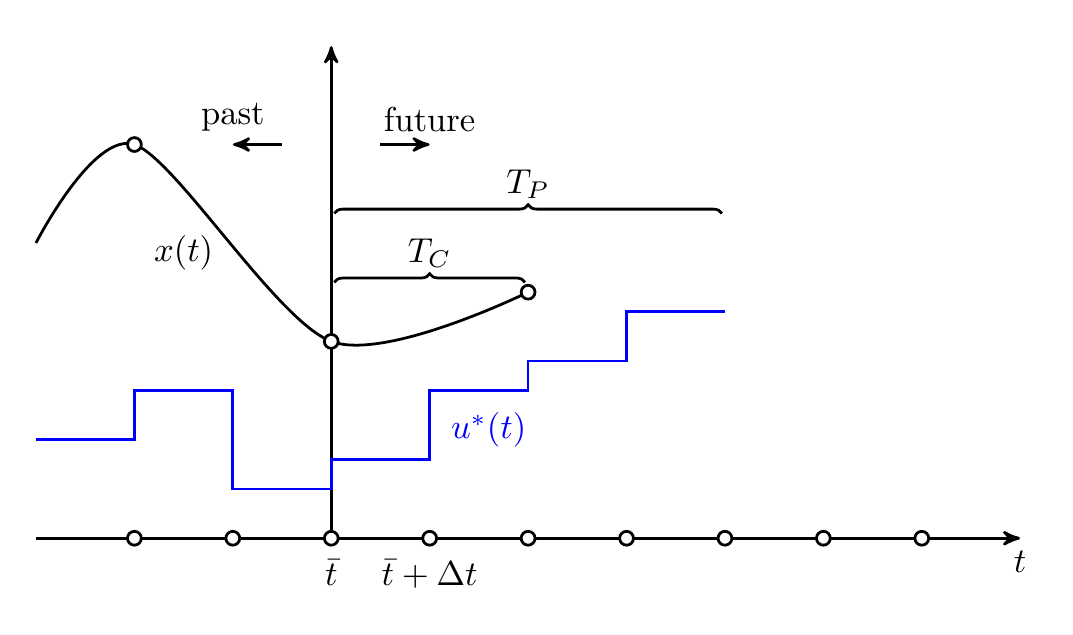} \label{mpc}
\caption{Schematic illustration of the MPC methodology. }
\end{center}
\end{figure}

 \begin{definition}
 We denote any agent who computes the exact Nash equilibrium solution of the system \eqref{microModelOpt} a \textbf{rational agent}. 
 \end{definition}
 This definition fits to the economic theory of rationality \cite{fama1965behavior, malkiel2003efficient}, so each agent is aware of the correct dynamic and acts fully optimal in the context of Nash equilibria. 
 We want to point out that in case of many agents, we have a large system of optimization problems \eqref{microModelOpt}. Such a system is very expensive to solve. 
 Thus, not only from the perspective of behavioral finance but also from a pure computational aspect such a rational setting seems to be very unrealistic. 
 Hence, we want to give a precise mathematical definition of so called \textbf{bounded rational agents} in the sense of Simon \cite{simon}. 
 \begin{definition}
We denote any agent who computes a numerical approximation of the microscopic system \eqref{microModelOpt} a \textbf{bounded rational agent}. 
 \end{definition}
We approximate the objective functional in \eqref{microModelOpt} by linear model-predictive control (MPC) \cite{michalska1989receding, camacho2013model}. 
This methods approximates a finite horizon optimal control problem in two ways. First, one predicts the dynamics over a predict horizon $T_P\leq T$ and secondly 
the control is only selected on a control horizon $T_C\leq T_P$. 
Thus for an arbitrary initial time $\bar{t}$ the optimization \eqref{microModelOpt} is performed on $[\bar{t},\bar{t}+T_P]$. Then the computed admissible control $\bar{u}: [\bar{t}, \bar{t}+T_P]\to \R$ can be applied 
on $[\bar{t},\bar{t}+T_C]$ and the state dynamics evolve accordingly on $[\bar{t},\bar{t}+T_C]$. Then the whole procedure is repeated with updated initial conditions at $\bar{t}_{new}:=\bar{t}+T_C$, shifted prediction interval $[\bar{t}_{new},\bar{t}_{new}+T_P]$ and control interval $[\bar{t}_{new}, \bar{t}_{new}+T_C]$. The algorithm terminates if the time $T$ is included in the prediction interval. 
A schematic illustration of one step of the linear MPC method is depicted in Figure \ref{mpc}. For sure, we can only expect to obtain a sub-optimal strategy compared to the original mode \eqref{microModelOpt}. This procedure can be considered as a repeated open-loop control in a feedback fashion. The computation on the prediction horizon is open-loop. Then one evolves the system on the control horizon and thus obtains a feedback by the system, which is used as initial condition for the next open-loop optimization.  
In order to apply MPC on our model we assume that $T_C=T_P$ holds. Furthermore, we discretize our dynamics on time intervals of length $\Delta t >0$ with $\ T=M\ \Delta t$ and we ensure that $\Delta t\leq T_C$ holds. 
Then the prediction and control horizon can be defined as $T_C=T_P=p \Delta t,\ 0<p\leq M$.

In fact, we assume that the optimal control $u_i^*$ is well approximated by piecewise constant controls of length $\Delta t$ on  $[\bar{t}, \bar{t}+M\ \Delta t]$.
$$
u_i^{*}(t)\approx \sum\limits_{k=0}^{M-1} u_i^k\ \mathds{1}_{[t_k,t_{k+1})}(t),\quad t_k=\bar{t}+k\ \Delta t,\quad k=0,...,M.  
$$

We choose the penalty parameter $\mu$ in the running costs to be proportional to the time interval so that $\mu = \nu \Delta t$ for some $\nu$. This can be motivated by checking the units of the variables in the cost functional ($K$ is a rate, thus measured in $1/\text{time}$, $\Psi$ is $\text{wealth}^2/\text{time}$, $u$ $\text{wealth}/\text{time}$). We see that the penalty parameter $\mu$ must be a time unit. Furthermore, we insert the right-hand side of the stock price equation into the stock return. Thus, the semi-discretized constrained optimization problem on $[\bar{t}, \bar{t}+p\ \Delta t]$ reads
\begin{align*}
&\int_{\bar{t}}^{\bar{t}+ p\Delta t} \left( \frac{\nu \Delta t}{2} u_i^2(t) + \Psi_i(t)\right) dt \to \min \\
&\dot x_i(t) = \kappa\ ED_N(t)\ x_i(t) + \frac{D(t)}{S(t)}\ x_i(t)+ u_i(t),\quad x_i(\bar{t}) =\bar{x}_i,\\
& \dot y_i(t) = ry_i(t) - u_i(t),\quad y_i(\bar{t})=\bar{y}_i,\\ 
&\dot{S}(t) = \kappa\ ED_N(t)\ S(t),\quad S(\bar{t}) = \bar{S}. 
\end{align*}

Thus, we can finally state a precise notion of bounded rationality in the context of MPC approximations.

\begin{definition}
Thanks to the MPC framework we can define the \textbf{degree of rationality} $\theta \in [0,1]$ of an agent by:
$$
\theta = 1-\exp\left\{\frac{1}{T}- \frac{T_P}{\Delta t\ T}\right\} ,\quad T_P=p\ \Delta t,\ p=1,...,\left \lfloor \frac{T}{\Delta t} \right\rfloor ,\ \text{and}\ T\gg \Delta t,
$$
where $\theta =1$ corresponds to a fully rational agent and  $\theta=0$ corresponds to a fully myopic agent in the MPC setting. 
Here, we assume that the model is at the beginning of the considered time period $t=0$. 
\end{definition}
In fact a rational agent is obtained for $\theta=1$, which is only the case if $T_P=T$ and $\Delta t\to 0$ holds. 
A fully myopic agent ($\theta=0$) is obtained for $T_P=\Delta t$. 
  
\begin{remark}
As the previous definition reveals, the MPC method introduces two different kinds of errors. A discretization error due to the numerical approximation of the optimal control and a prediction error due to the approximated control horizon. There are many contributions which derive perfomance bounds of the MPC method in order to quantify the impact of a limited time horizon \cite{grune2008infinite}. 
The impact of time varying control horizons $T_C$ on the performance of the MPC method in comparison to the exact closed-loop solution has been discussed by Gr\"une et al. \cite{grune2010analysis}. 
\end{remark}

\paragraph{Optimality system}
As pointed out previously, we want to solve the MPC problem in a game theoretic setting. We want to search for Nash equilibria. In this setting, each agent assumes that the strategies of the other players are fixed and optimal. Thus, we get $N$ optimization problems which need to be solved simultaneously. Hence, we have a $N$-dimensional Lagrangian $L\in\R^N$.  The i-th entry $L_i$ corresponds to the i-th player and reads:

\begin{align}
L_i(x_i,y_i, S, u_i, \lambda_{x_i}, \lambda_{y_i}, \lambda_{S})=
&\int\limits_{\bar{t}}^{\bar{t}+p\ \Delta t}\left( \frac{\nu \Delta t}{2} u_i^2(t) + \Psi_i(t)\right) dt\label{lagrange} \\
&+\int\limits_{\bar{t}}^{\bar{t}+p\ \Delta t}\dot{\lambda}_{x_i}\ x_i+ \lambda_{x_i}\ \kappa\ ED_N \ x_i + \lambda_{x_i}\ \frac{D}{S} \ x_i +\lambda_{x_i}\ u_i\ dt -\lambda_{x_i}\ \bar{x}_i\\
&+\int\limits_{\bar{t}}^{\bar{t}+p\ \Delta t}\dot{\lambda}_{y_i}\ y_i+ \lambda_{y_i}\ r\ y_i -\lambda_{y_i}\ u_i\ dt-\lambda_{y_i}\ \bar{y}_i,\\
&+\int\limits_{\bar{t}}^{\bar{t}+p\ \Delta t}\dot{\lambda}_{S}\  S+ \lambda_{S}\ \kappa\ ED_N\ S \ dt-\lambda_{S}\ \bar{S},
\end{align}  
with Lagrange multiplier $\lambda_{x_i}, \lambda_{y_i}, \lambda_{S} $. 
Notice that the quantities $(x^*_j,y^*_j,u^*_j),\ j=1,...,i-1,i+1,...,N $ are assumed to be optimal in the i-th optimization and therefore only enter as parameters in the i-th Lagrangian $L_i$. We assume $\lambda_{x_i}(\bar{t}+p\Delta t)= \lambda_{y_i}(\bar{t}+p\Delta t)= \lambda_{S}(\bar{t}+p\Delta t)=0$. The optimal control can be obtained by solving the corresponding necessary optimality conditions. 
The number of optimality conditions depends on the size of the prediction horizon $T_P=p\ \Delta t$. Thus, for each agent one needs to solve an optimality system consisting of $n=p\ 7$ equations. 
For further details on differential games we refer to \cite{bressan2011noncooperative}.\\ \\
The previously introduced framework for the degree of rationality of investors in the context of portfolio optimization is rather general. Therefore we will specify the running cost in the following paragraph.

\paragraph{Fully myopic agent}
In the further discussion of our model we confine the study on the case $p=1$. This choice has several reasons: First of all this simplification enables us to compute the optimal control explicitly and thus to derive the macroscopic limit of our microscopic dynamics. As a direct consequence, we obtain a three dimensional ODE system which we are able to analyze analytically. Finally, the simulations of our resulting ODE model reveals that our model shares many similarities with macroeconomic financial market models in literature. \\ \\
In the case of a fully myopic agent ($\theta=0$), which corresponds to $p=1$ we can even compute the optimal control explicitly. We still have to define our objective function $\Psi_i$ that determines the agent's actions. We assume that the agent minimizes a quantity proportional to the expected missed revenues in each portfolio. 
The quantity $K$, which we define in detail later, is a return estimate of the stock return over the bond return. One can expect that $K$ depends on the current or past stock prices. If stocks are believed to be better ($K>0$), then being invested in bonds ($y_i>0$) is bad, and vice versa. Then $|K|\ x_i$ for $K<0$ is the expected missed revenue of agent $i$ by having invested in stocks but not in bonds. Equivalently, $|K|\ y_i$ for $K>0$ is the expected missed revenue of agent $i$ by having invested in bonds but not in stocks.
Then we weight the expected missed revenue by the wealth in the corresponding portfolio and define the running cost by
\begin{align*}
\Psi_i := 
\begin{cases}
 |K|\ \frac{x_i^2}{2},&\ K<0,\\
  0,&\ K=0,\\
|K|\ \frac{y_i^2}{2},&\ K>0.
\end{cases}
\end{align*}
The weighed missed revenue is larger, the larger the estimated difference between returns $K$. We still need to define the precise shape of the return estimate $ K$. 
\\ \\
As done in many asset pricing models, we consider a chartist and a fundamental trading strategy. A trading strategy refers to an estimate of future stock return in order to evaluate the profit of the portfolio. \\  \\
Fundamentalists believe in a fundamental value of the stock price denoted by $s^f>0$ and assume that the stock price will converge in the future to this specific value. The investor therefore estimates the future return of stocks versus the return of bonds as
\begin{align*}
K^f:= U_{\gamma}\left( \omega \frac{s^f-S}{S}\right)-r.
\end{align*}
Here, $U_{\gamma}$ is a value function in the sense of Kahnemann and Tversky \cite{kahneman1979prospect} which depends on the risk tolerance $\gamma$ of an investor. A typical example is $U_\gamma(x) = sgn(x) |x|^{\gamma}$ with $0<\gamma<1$ and sign function $sgn$.
The constant $\omega>0$ measures the expected speed of mean reversion to the fundamental value $s^f$. We want to point out that this stock return estimate is a rate and thus $\omega$ needs to scale with time.\\
Chartists assume that the future stock return is best approximated by the current or past stock return. They estimate the return rate of stocks over bonds by
\begin{align*}
{K}^c:=U_{\gamma}\left( \frac{\dot{S}+D}{S} \right)-r.
\end{align*}
Both estimates are aggregated into one estimate of stock return over bond return by a convex combination
$$
K = \chi\  K^f + (1-\chi)\ K^c.
$$
This idea has been previously applied to a kinetic model of opinion formation \cite{albi2014boltzmann}. The weight $\chi$ is determined from an instantaneous comparison as modeled in \cite{lux1999scaling}. We let
$$
\chi = W(K^f-K^c),
$$
where $W:\R\to [0,1]$ is a continuous function. If for example, $W = \frac12\text{tanh}+\frac12$, the investor optimistically believes in the higher estimate.
Together, if $K>0$, the investor believes that stocks will perform better and if $K<0$ that bonds will perform better. \\ \\
Thus, the necessary optimality conditions we obtain from our Lagrangian \eqref{lagrange} are given by
\begin{align*}
&\dot x(t) = \kappa\ ED_N(t)\ x_i(t) + \frac{D(t)}{S(t)}\ x_i+ u_i,\  x_i(\bar{t}) =\bar{x}_i,    \\
& \dot y_i(t) = ry_i(t) - u_i(t),\  y_i(\bar{t})=\bar{y}_i,   \\ 
& \dot{S}(t) = \kappa\ ED_N(t)\ S(t),\ S(\bar{t}) = \bar{S}, \\
&\nu\ \Delta t\ u_i(t) = - \lambda_{x_i}(t)- \lambda_{x_i}(t)\ \frac{\kappa }{N}\ x_i(t)  + \lambda_{y_i}(t)- \frac{\kappa}{N}\ S(t)\ \lambda_{S}(t),\\
& \dot{\lambda}_{x_i}(t) = -\kappa\ ED_N(t)\ \lambda_{x_i}(t)- \frac{D(t)}{S(t)}\ \lambda_{x_i}(t)- \partial_{x_i} \Psi_i(t),\ \lambda_{x_i}(\bar{t}+\Delta t)=0\\
& \dot{\lambda}_{y_i}(t) = -r \lambda_{y_i}(t)- \partial_{y_i} \Psi_i(t),\ \lambda_{y_i}(\bar{t}+\Delta t)=0\\
& \dot{\lambda}_{S}(t) =  \lambda_{x_i}(t)\frac{D(t)}{S^2(t)} x_i-\kappa\ ED_N(t)\ \lambda_{S}(t)- \partial_{S} \Psi_i(t),\ \lambda_{S}(\bar{t}+\Delta t)=0.
\end{align*}
Here, we neglect for the purpose of readability the dependence of the $ED_N(t)=ED_N(t,x_i,y_i, S)$ and $\Psi_i(t)= \Psi_i(t,x_i,y_i, S)$ on the wealth and stock price. 
Then we apply a backward Euler discretization to the adjoint equations and get.
\begin{align*}
&\frac{\lambda_{x_i}(t+\Delta t)-\lambda_{x_i}(\bar{t})}{\Delta t}= -\kappa\ ED_N(t+\Delta t)\ \lambda_{x_i}(t+\Delta t)- \frac{D(t+\Delta t)}{S(t+\Delta t)}\ \lambda_{x_i}(t+\Delta t)- \partial_{x_i} \Psi_i(t+\Delta t),\\
&  \frac{\lambda_{y_i}(t+\Delta t)-\lambda_{y_i}(\bar{t})}{\Delta t}= -r \lambda_{y_i}(t+\Delta t)- \partial_{y_i} \Psi_i(t+\Delta t),\\
& \frac{\lambda_{S}(t+\Delta t)-\lambda_{S}(\bar{t})}{\Delta t} =  \lambda_{x_i}(t+\Delta t)\frac{D(t+\Delta t)}{S^2(t+\Delta t)} x_i-\kappa\ ED_N(t+\Delta t)\ \lambda_{S}(t+\Delta t)- \partial_{S} \Psi_i(t+\Delta t).
\end{align*}
Then, we insert the final conditions of the costates and obtain.
\begin{align*}
&\lambda_{x_i}(\bar{t}) = \Delta t\ \partial_{x_i} \Psi_i(\bar{t}+\Delta t),\\
&\lambda_{y_i}(\bar{t}) = \Delta t\ \partial_{y_i} \Psi_i(\bar{t}+\Delta t),\\
&\lambda_{S}(\bar{t}) = \Delta t\ \partial_{S} \Psi_i(\bar{t}+\Delta t). 
\end{align*}
Hence, the optimal strategy is given by
\begin{align*}
u_N^*(x_i,y_i, S)= 
\begin{cases}
\frac{1}{\nu} ( {K}\ y_i - \frac{\kappa}{N}\ S\  (\partial_{S} {K})\ \frac{y_i^2}{2}),&\quad K>0,\\
0,&\quad K=0,\\
\frac{1}{\nu} ( {K}\ x_i + {K}\ \frac{\kappa}{N}\ x_i^2 + \frac{\kappa}{N} S\  (\partial_{S} {K})\ \frac{x_i^2}{2}),&\quad K<0. 
\end{cases}
\end{align*}
\paragraph{Instantaneous controlled model}
Thank to our simplified setting we could compute the control explicitly. In the engineering literature such a control is frequently called instantaneous control and our model reads:
\begin{subequations}
\label{eq:wealthN}
\begin{align}
& \dot{x}_i(t) =\kappa\ ED_N(t)\ x_i(t)+ \frac{D(t)}{S(t)}\ x_i(t)+u_N^*(t,x_i,y_i,S)\\
& \dot{y}_i(t) = r\ y_i(t)-u_N^*(t,x_i,y_i, S) \\
& \dot{S}(t) = \kappa\ ED_N(t)\ S(t).
\end{align}
\end{subequations}
Here, we have inserted the right-hand side of our stock equation \eqref{MacroStockODE} into the stock return.

\section{Macroscopic Portfolio Model}

In order to derive the macroscopic portfolio model we average the microscopic quantities and consider the limit of infinitely many agents.
This limit of infinitely many particles is known in physics as mean field limit \cite{stanley1971phase, golse2016dynamics}. We define the average wealth invested in stocks, respectively bonds by:
$$
X_N(t):= \frac{1}{N} \sum\limits_{i=1}^N x_i(t),\quad Y_N(t):= \frac{1}{N} \sum\limits_{i=1}^N y_i(t).
$$
Furthermore, we assume that the limits 
$$
\lim\limits_{N\to\infty} X_N(t) = X(t)<\infty ,\quad  \lim\limits_{N\to\infty} Y_N(t) = Y(t)<\infty,
$$
exist for all times. The only non-linearity in $x_i, y_i$ is the average of investment strategies.  This is nothing else than our excess demand we considered before
\[  
ED_N= \frac{1}{N} \sum\limits_{i=1}^N u_i^*=\begin{cases}
\frac{1}{\nu} \Big( {K}\frac{1}{N}\sum\limits_{i=1}^N\ y_i - \frac{\kappa}{N}\ S\  (\partial_{S} {K})\ \frac{1}{2\ N}\sum\limits_{i=1}^N y_i^2\Big), &\quad K>0,\\
0,&\quad K=0,\\
\frac{1}{\nu} \Big( {K}\ x_i + {K}\ \frac{\kappa}{N}\ \frac{1}{N}\sum\limits_{i=1}^N x_i^2 + \frac{\kappa}{N} S\  (\partial_{S} {K})\ \frac{1}{2\ N}\sum\limits_{i=1}^N  x_i^2\Big),&\quad K<0. 
\end{cases} 
\]
In the limit of infinitely many agents we obtain
\begin{align*}
\lim\limits_{N\to\infty}ED_N=ED(t,X,Y,S,\dot{S})&:= 
\begin{cases}
\frac{1}{\nu}K(t,S,\dot{S})\ X(t),&\quad K(t,S,\dot{S})<0,\\
0,&\quad K(t,S,\dot{S})=0,\\
\frac{1}{\nu}K(t,S,\dot{S})\ Y(t),&\quad K(t,S,\dot{S})>0.
\end{cases}
\end{align*}
Here, we have used that the quadratic terms
$$
 \frac{1}{N^2} \sum\limits_{i=1}^N x_i^2(t),\quad \frac{1}{N^2} \sum\limits_{i=1}^N y_i^2(t),
$$
vanish in the limit, which can be made evidend by analyzing the order of:
$$
\infty> \left( \frac{1}{N}\sum\limits_{i=1}^N x_i \right)^2 = \frac{1}{N^2} \sum\limits_{i=1}^N x_i^2+ \frac{1}{N^2}  \sum\limits_{1\leq k,j \leq N,\  k\neq j} x_k\ x_j.
$$ 
\begin{remark}
A rigorous derivation of the macroscopic portfolio model can be performed by the use of mean field theory. We refer to \cite{trimborn2017portfolio} for details. 
\end{remark}

\paragraph{Macroscopic portfolio model}
Then in order to obtain the limit equation we simply sum over the number of agents and consider the limit $N\to\infty$. 
Thus, the macroscopic portfolio model reads.
 \begin{subequations}\label{moment}
 \begin{align}
&\frac{d}{dt} X(t) =  \frac{\dot{ S}(t)+D(t)}{S(t)}\ X(t)+ ED(t,X,Y,S,\dot{S})\\
&\frac{d}{dt} Y(t) = r\ Y(t) - ED(t,X,Y,S, \dot{S})\\
&\frac{d}{dt} S(t) = \kappa\  ED(t,X,Y,S, \dot{S})\  S(t).
\end{align}
 \end{subequations} 
 The macroscopic wealth variable $W(t)>0$ is given by $W(t)=X(t)+Y(t),\ X(t),Y(t)>0$ and can be regarded as the average wealth in a society or country.
  Hence, we have a three dimensional ODE system in which the stock price equation is an implicit ODE.

\section{Qualitative Analysis }

This section is devoted to analyze the rich dynamics of the ODE system \eqref{moment}. We discuss existence and uniqueness of solution, steady states and their stability and even compute explicit solutions in special cases.

\paragraph{Macroscopic steady states} In order to obtain steady states, the equations 
\begin{align*}
0= \kappa\ ED(X,Y,S)X+ \frac{D}{S} X+ ED(X,Y,S)\\
0= rY - ED(X,Y,S)\\
0= \kappa\  ED(X,Y,S)\  S,
\end{align*}
need to be fulfilled. Besides the trivial solution the following steady state configurations are possible.
\begin{itemize}
\item[i)] $X=0$,\quad $Y=0$,\quad $S$ arbitrary
\item[ii)] $K(S)=0$,\quad $Y=0$,\quad $D=0$,\quad $X$ arbitrary
\item[iii)] $K(S)=0$,\quad $r=0$,\quad $D=0$,\quad $X$ and $Y$ arbitrary
\item[iv)] $K(S)>0$,\quad $Y=0$,\quad $D=0$,\quad $X$ arbitrary
\item[v)] $K(S)<0$,\quad $X=0$,\quad $r=0$,\quad $Y$ arbitrary
\end{itemize} 
The case $i)$ corresponds to the situation when all investors are bankrupt. 
In the cases $ii)$ and $iii)$, the investors expect to have no benefit of shifting the capital between both portfolios.
This means that the expected return $K(S)$ is zero, which is equivalent to
$$
U_{\gamma}\left(\omega\ \frac{s^f-S}{S}\right)\ \chi+ (1-\chi)\ U_{\gamma}\left( 0\right)=r.
$$
If we choose the value function $U_{\gamma}$ to be the identity, we observe 
$$
S^{\infty} = \frac{\chi\ \omega\ s^f}{1+r},
$$
as the equilibrium stock price. One might assume that the reference point of the value function is not zero. This means that the financial agent has a fixed bias towards potential gains or losses. Mathematically, $U_{\gamma}(0)\neq 0$ holds and thus the steady state would be shifted by the reference point. Hence, psychological misperceptions of investors lead to changes of the equilibrium price. The case $iv)$ corresponds to the situation that the investor wants to shift wealth from the bond portfolio into the stock portfolio. In fact, no transaction takes place, since there is no wealth left in the bond portfolio. Thus $K(S)>0$ has to hold, which means
$$
U_{\gamma}\left(\omega\ \frac{s^f-S}{S}\right)\ \chi+ (1-\chi)\ U_{\gamma}\left( 0\right)>r.
$$
In the simple case of the identity function as utility function, we obtain:
\begin{align}
\frac{\omega\ \chi\ s^f}{r+\omega\ \chi}> S .\label{inequalitySteady}
\end{align}
In this equilibrium case, the amount of transactions have been too low to push the price above a certain threshold defined by inequality \eqref{inequalitySteady}. 
The reason for the steady state is the bankruptcy in the bond portfolio. Such a situation does not reflect a usual situation in financial markets. 
In case $v)$, we face the opposite situation. Here, the investor wants to shift wealth from stocks to bonds although there is no wealth left in the stock portfolio. 

\subsection{Simplified Model}

In order to give a detailed characterizations of the complex dynamics we assume during the rest of the section that the weight $W\in [0,1]$ is constant and that the value function $U_{\gamma}$ is given by the identity. In this setting it is possible to obtain local Lipschitz continuity directly. 

\begin{proposition}\label{prop2}
With the previously stated assumptions we can ensure existence and uniqueness of a solution (at least for short times $[t_0, t_0 +\epsilon]$).
\end{proposition}
In addition, we are interested in the stability of the steady state $S^{\infty} = \frac{\chi\ \omega\ s^f}{1+r}$ characterized in $iii)$.
From economic perspective this is the only reasonable stationary point.

\begin{proposition}\label{prop3}
In addition to the previously stated assumptions we assume that $D=r=0$ holds. Then $S^{\infty} = \chi\ \omega\ s^f$ is a unique asymptotically stable steady state. 
\end{proposition}

For the proof of both  results we refer to the appendix. Furthermore, we state the explicit solutions of the stock price and portfolio dynamics in the appendix \ref{compMom}. 
In the subsequent paragraph a qualitative discussion of the observed dynamics is given.

\paragraph{Booms, crashes and oscillatory solutions}
We want to study whether the stock price satisfies the most prominent features of stock markets. These are crashes, booms and oscillatory solutions. 
Mathematically, a boom or crash is described by exponential growth or decay of the price.   
\begin{itemize}
\item Fundamentalists merely ($\chi=1$) influence the price by their fundamental value $s^f$. The price is driven to the steady state $S_{\infty}=\frac{\omega\ s^f}{\omega+r}$ exponentially. Interestingly, the convergence speed depends on the market depth $\kappa$, the interest rate $r$, the expected speed of mean reversion $\omega$ and the amount of wealth invested.
\item Chartists merely ($\chi = 0$) build their investment decision on the current stock return. 
The price gets driven exponentially to the equilibrium stock price $S_{\infty}=\frac{D}{r}$ or away from the equilibrium stock price. This behavior is determined by the average wealth invested in stocks or bonds. In general, we observe exponential growth or decay of the stock price (e.g. $D\equiv 0$).  Hence, the chartist behavior can create market booms or crashes. We can thus expect that an interplay of fundamental and chartist strategies leads to oscillatory behavior around the equilibrium prices.
\item In our last case, we consider a mix of chartist and fundamental return expectations with a constant weight $\chi\in (0,1)$. 
In that case, the price converges to the equilibrium price $S_{\infty}= \frac{\chi\ \omega \ s^f+(1-\chi)\ D}{\chi\ \omega +r}$
which is a combination of the previous equilibrium prices. Thus, the weight $\chi$ heavily influences the price dynamic.
Furthermore, we can expect to observe oscillatory solutions if we consider a non constant weight $\chi(t,S)$. 
\end{itemize}

\paragraph{Wealth evolution} We can analyze the wealth evolution in the same manner as previously the stock price equation. We consider each portfolio separately. The computation can be found in the appendix \ref{compMom}, as well.
\begin{itemize}
\item We have exponential growth in the stock portfolio, if the wealth gets transferred from bonds to stocks.
In the opposite case, the decay of wealth is described by an exponential as well. 
\item In the bond portfolio, we also observe an exponential increase if the wealth gets shifted into the bond portfolio. 
If stocks are assumed to perform substantially better ($K(S)>r)$), we have exponential decay in the bond portfolio.
\end{itemize}

So far we have only discussed the simplified case of a constant weight and the identity function as value function. The previous discussion indicates that a nonlinear interplay of a fundamental and chartist strategy may cause oscillatory behavior. A rigorous quantification of this behavior is left open for further research.

 \section{Simulations}
 We want to provide insights into the portfolio dynamics of the model. Furthermore, we intend to determine the influence of each parameter in the model.
 We will verify the existence of oscillatory solutions of the model.  For simulations we choose the value function $U_{\gamma}$ and the weight function $W$ as follows:
 \begin{align*}
 &W(K^f-K^c):= \beta\ \left(\frac{1}{2} \tanh\left(\frac{K^f-K^c}{\alpha}\right)+\frac12\right) +(1-\beta)\ \left(\frac{1}{2} \tanh\left(-\frac{K^f-K^c}{\alpha}\right)+\frac12\right),\\
 &\quad\quad  \alpha>0,\  \beta\in[0,1],\\
 &
 U_{\gamma}(x):= \begin{cases}
 x^{\gamma+0.05},\quad  x>0,\\
 -(|x|)^{\gamma-0.05},\quad x\leq 0,\quad \gamma\in [0.05,0.95].
 \end{cases}
 \end{align*}
The weight function $W$ models the instantaneous comparison of the fundamental and chartist return estimate. 
The constant $\beta\in [0,1]$ determines if the investor trusts in the higher ($\beta=1$) or lower estimate ($\beta=0$) and we thus call this constant the trust coefficient. The constant $\alpha>0$ simply scales the estimated returns.

\begin{figure}[h!]
\begin{center}
\includegraphics[scale=0.3]{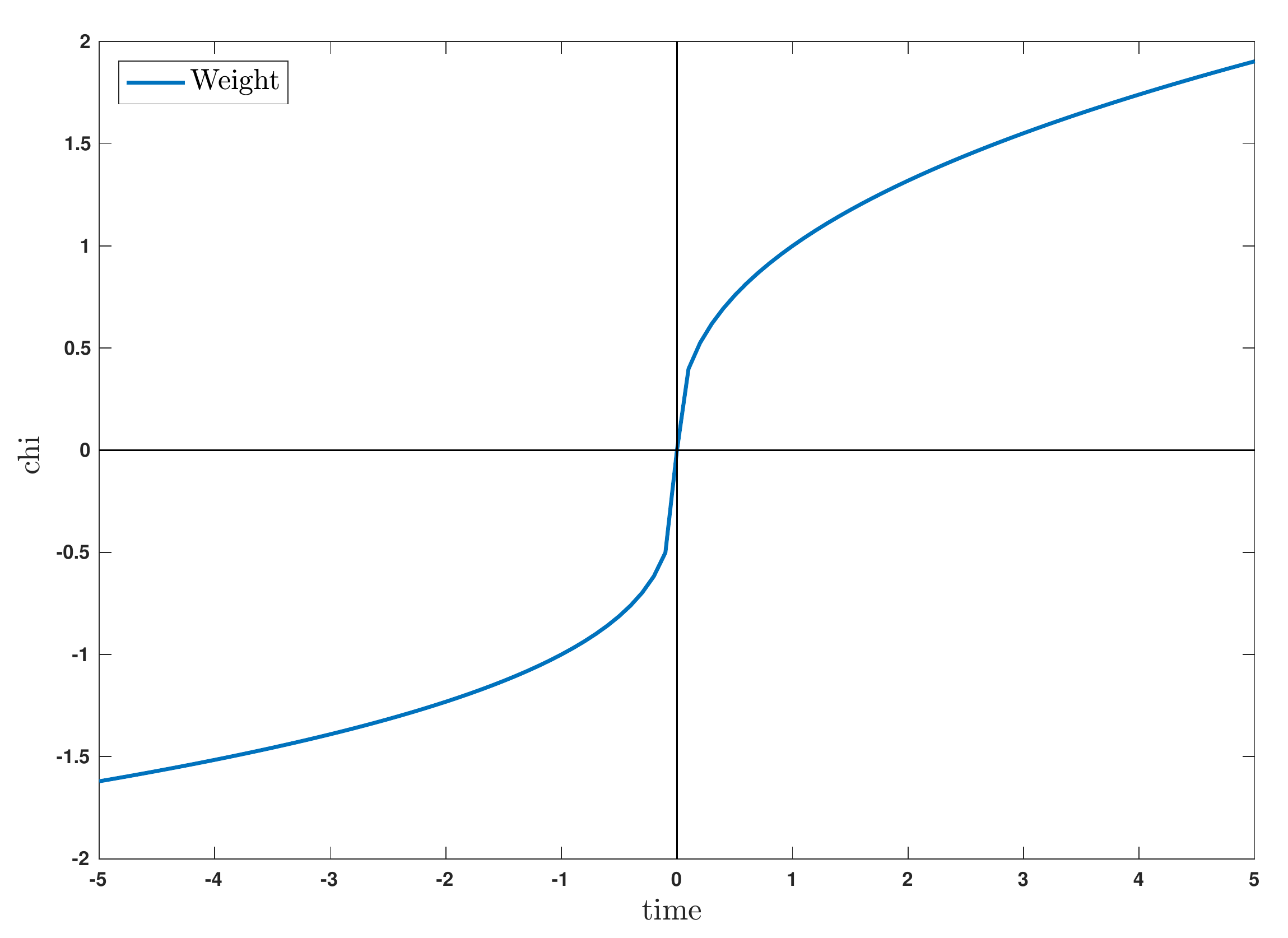}
\hfill
\includegraphics[scale=0.3]{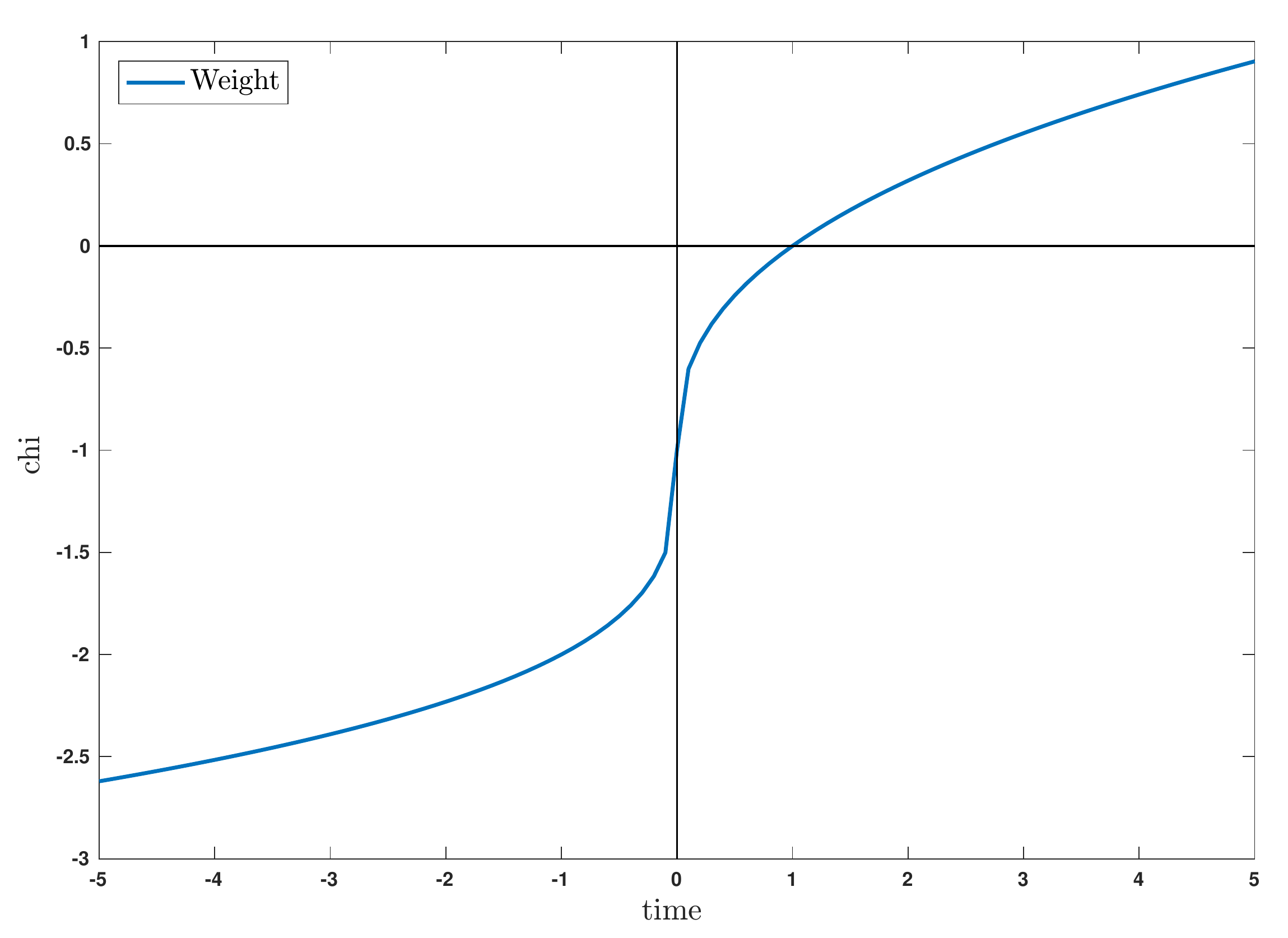}
\caption{Example of Value functions with different reference points.}\label{ValueFunction}
\end{center}
\end{figure}
\begin{figure}[h!]
\begin{center}
\includegraphics[scale=0.31]{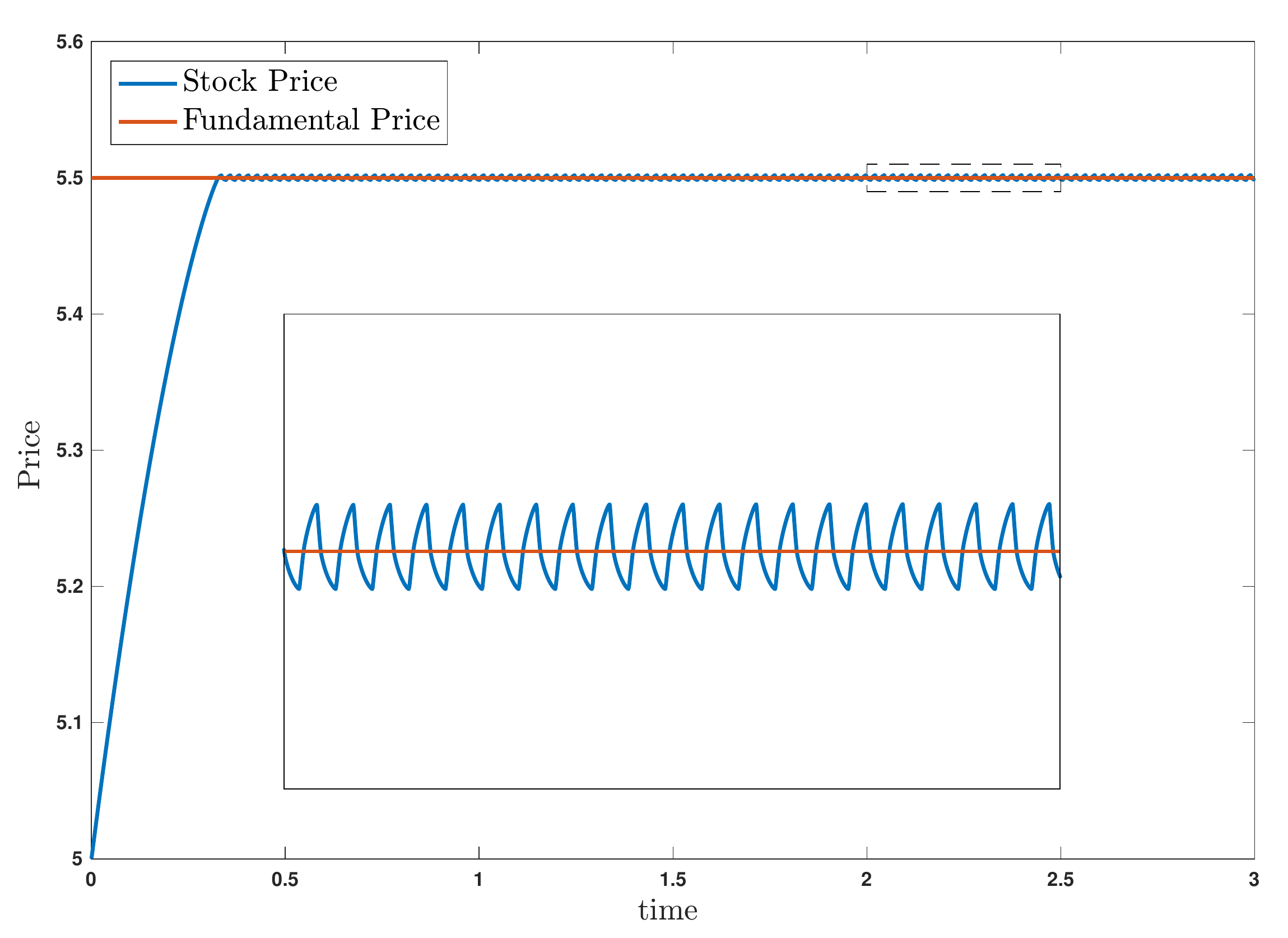} 
\hfill
\includegraphics[scale=0.31]{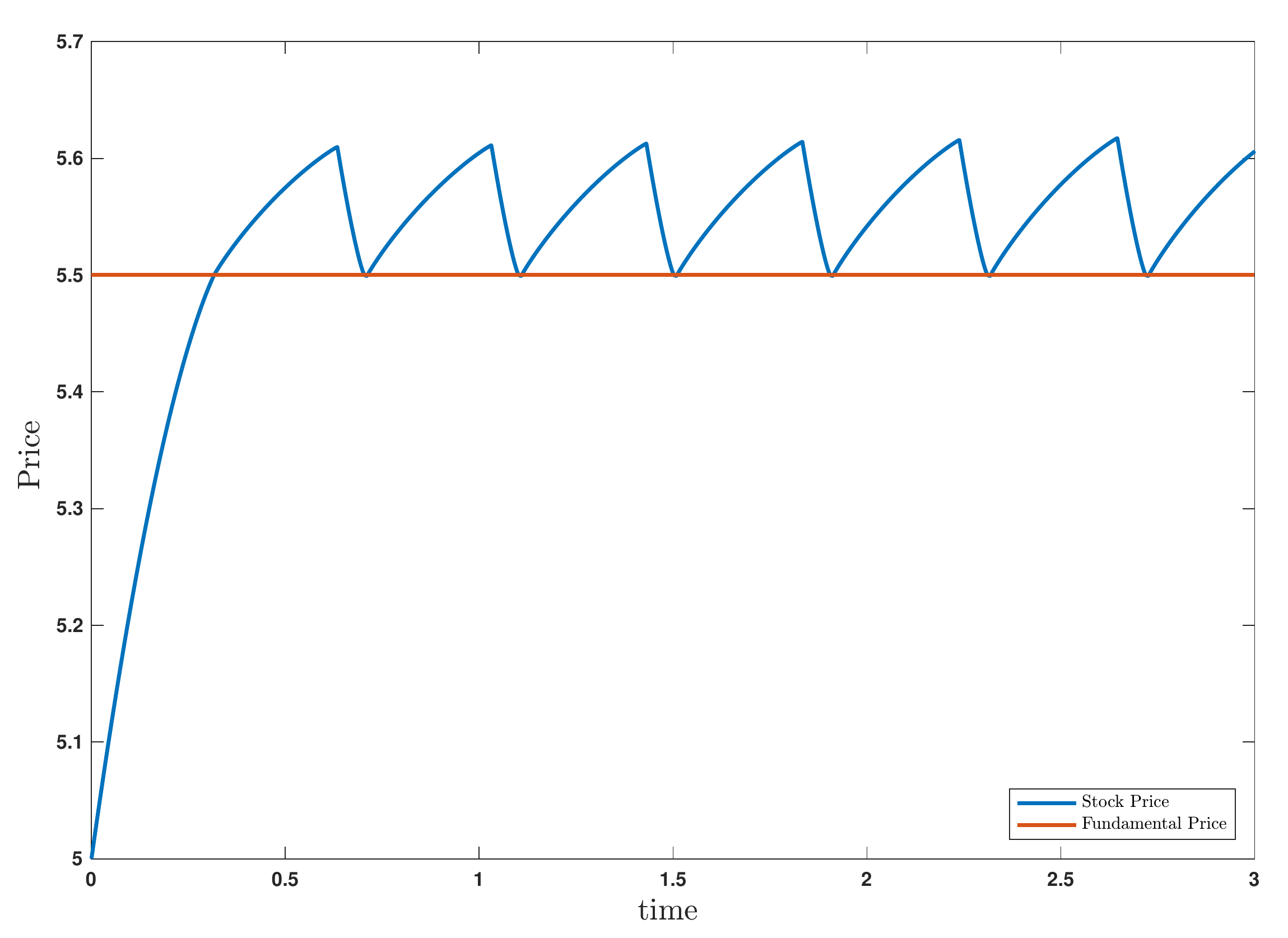} 
\end{center}
\caption{Stock price with trust coefficient $\beta=0.7$ (left figure) and trust coefficient $\beta = 0.85$ (right figure).}\label{alterTrust}
\end{figure}

The value function $U_{\gamma}$ models psychological behavior of an investor towards gains and losses. In order to derive the value function, one needs to measure the attitude of an individual as a deviation from a reference point. We have chosen the reference point to be zero, since $U_{\gamma}(0)=0$ holds. In Figure \ref{ValueFunction} we have plotted $U_{\gamma}$ and $\bar{U}_{\gamma}:= U_{\gamma}-1$ . The value function $\bar{U}_{\gamma}$ is an example of a value function with a negative reference point. This choice of value function satisfies the usual assumptions: the function is concave for gains and convex for losses, which corresponds to risk aversion and risk seeking behavior of investors. Furthermore, the value function is steeper for losses than for gains, which models the psychological loss aversion of financial agents (see Figure \ref{ValueFunction}).\\ [2em]
We have solved the moment system with a simple forward Euler discretization. The time step has been chosen sufficiently small to exclude stability problems due to stiffness. We verified the results with the \texttt{ode15i} Matlab solver. In fact one time step $\Delta t$ may correspond to one trading day. Then our simulation with $30,000$ time steps may correspond to approximately $120$ years of trading.\\ 
\begin{figure}[h!]
\begin{center}
\includegraphics[scale=0.5]{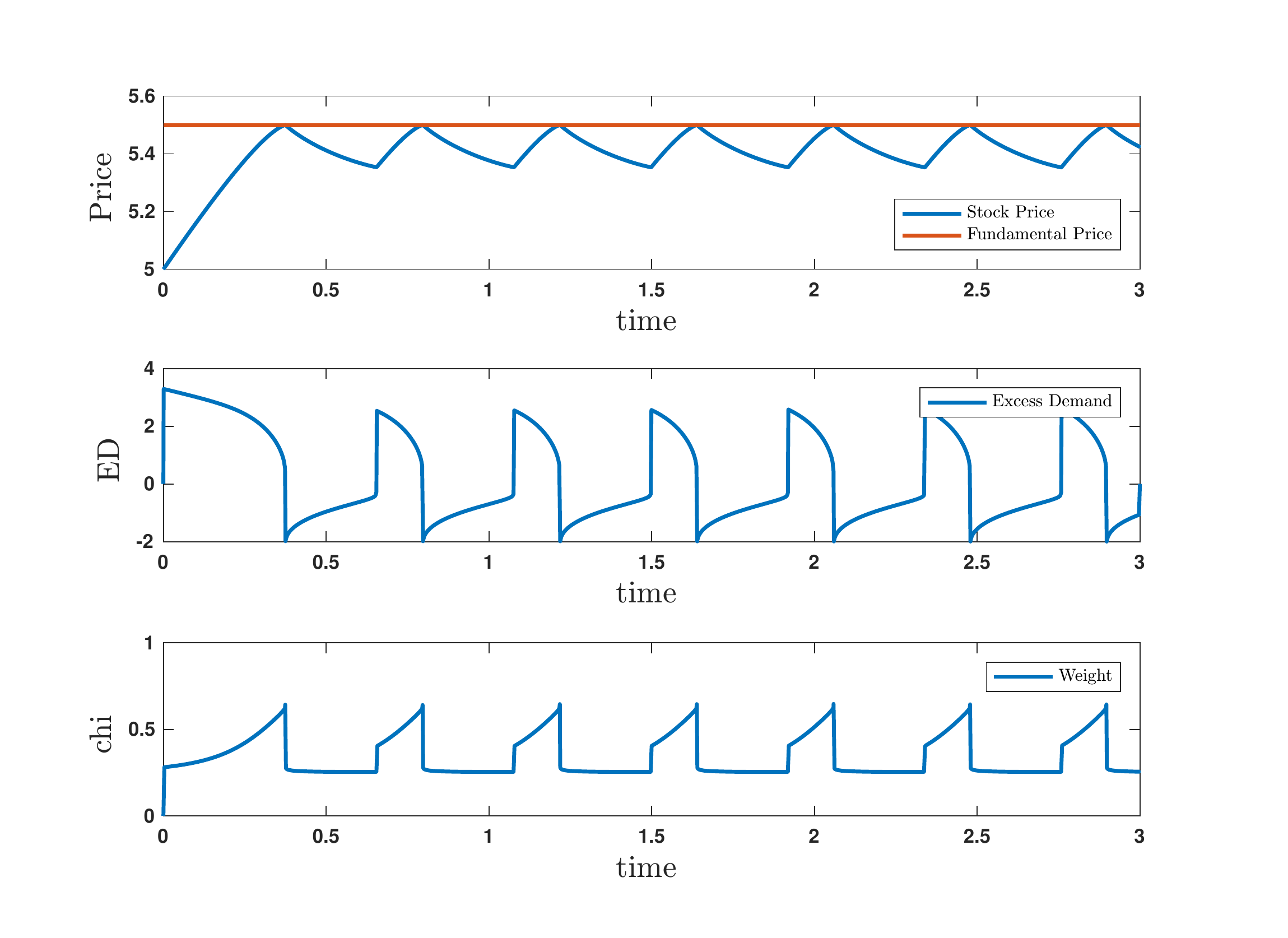}
\end{center}
\caption{Stock price, excess demand and weight $\chi$.}\label{momentstock}
\end{figure}
\begin{figure}[h!]
\begin{center}
\includegraphics[scale=0.5]{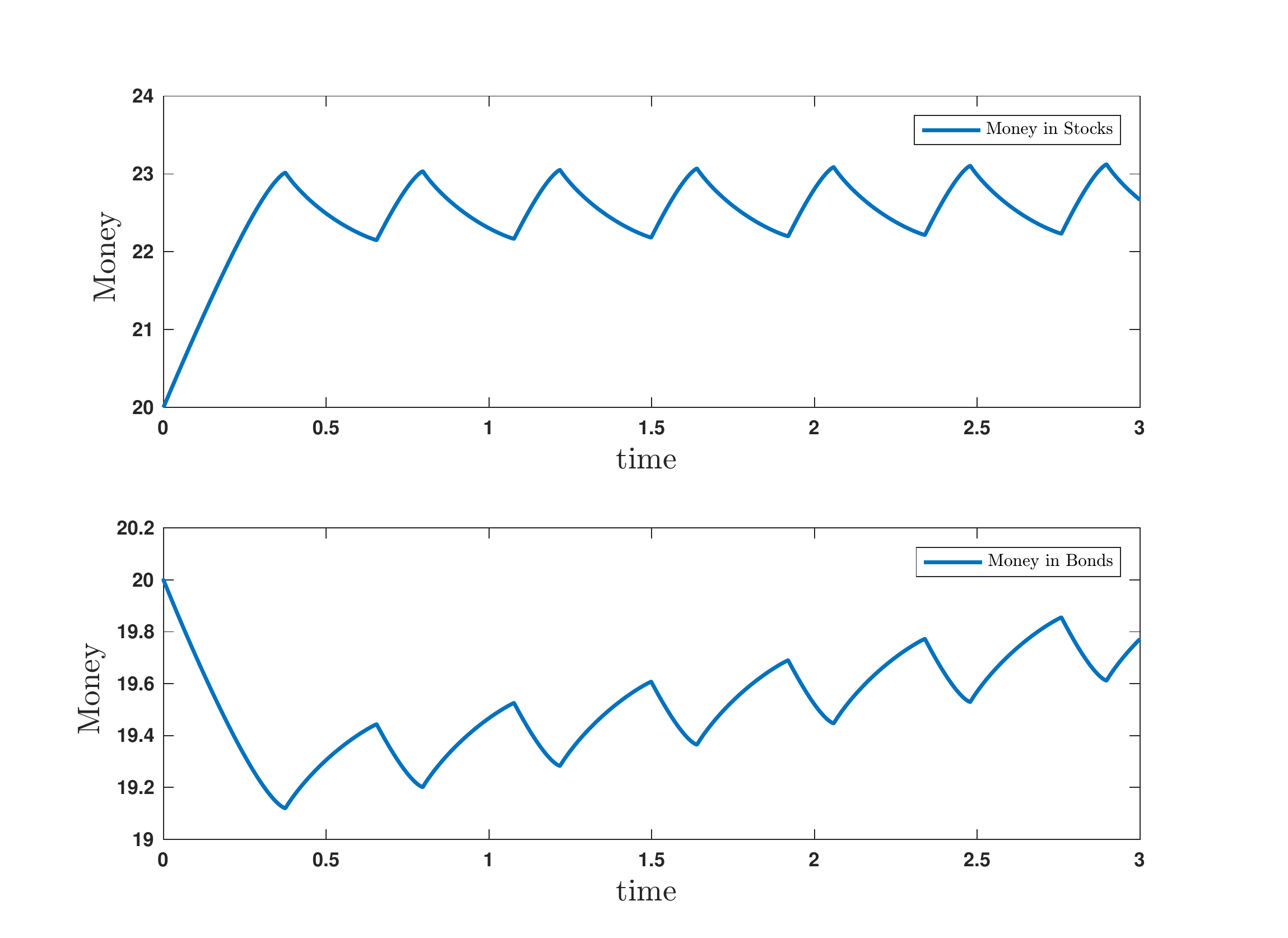}
\end{center}
\caption{Wealth evolution.}\label{wealthmoment}
\end{figure}
We have chosen a trust coefficient $\beta=0.25$ for the simulations in Figure \ref{momentstock} and \ref{wealthmoment}. We refer to the appendix \ref{paramMom} for further settings.
 The oscillations of the stock price is caused by oscillations in the excess demand. The stock price is always less than or equal to the fundamental price.
In addition, the oscillations get translated to the wealth evolution of the portfolios. Increasing wealth in the stock portfolio leads to decreasing wealth in the bond portfolio. 
Furthermore, we can observe on average a small positive slope of the wealth invested in bonds (see Figure \ref{wealthmoment}). This is caused by the positive interest rate $r$.  
In the next simulations (Figure \ref{alterTrust}), we have altered the trust coefficient to study the impact on the price behavior. 
As Figure \ref{alterTrust} reveals, the trust coefficient $\beta$ influences the amplitude and frequency of the oscillations.
In fact the oscillations obtained for the values $\beta=0.25$ and $\beta=0.85$ can be interpreted as business cycles since they last approximately for $10$ years. 
In addition, $\beta$ determines the location of the oscillatory stock price evolution with respect to the fundamental value $s^f$. A low trust coefficient leads to oscillations located below the fundamental price and a high trust coefficient to oscillations above the fundamental value.\clearpage
We want to point out that the price behavior is very sensitive with respect to the parameters $\gamma, \kappa, \omega, \alpha$ and $\beta$.
\begin{remark}\label{remark}
The parameters influence the price dynamics as follows:
\begin{itemize}
\item A larger risk tolerance $\gamma$ leads to smaller wave periods and smaller amplitudes. 
      A high risk tolerance heavily changes the price characteristics. We could thus observe convergence of the price to the fundamental value.  
\item The market depth $\kappa$ influences the amplitude of the oscillations. A bigger $\kappa$ value leads to a larger amplitude.  
\item The speed of mean reversion $\omega$, the scale parameter $\alpha$ influence the wave period and amplitude.
The wave period and amplitude decrease with increasing $\omega$, respectively $\alpha$. 
\end{itemize}
\end{remark}

\begin{figure}[h!]
\begin{center}
\includegraphics[scale=0.3]{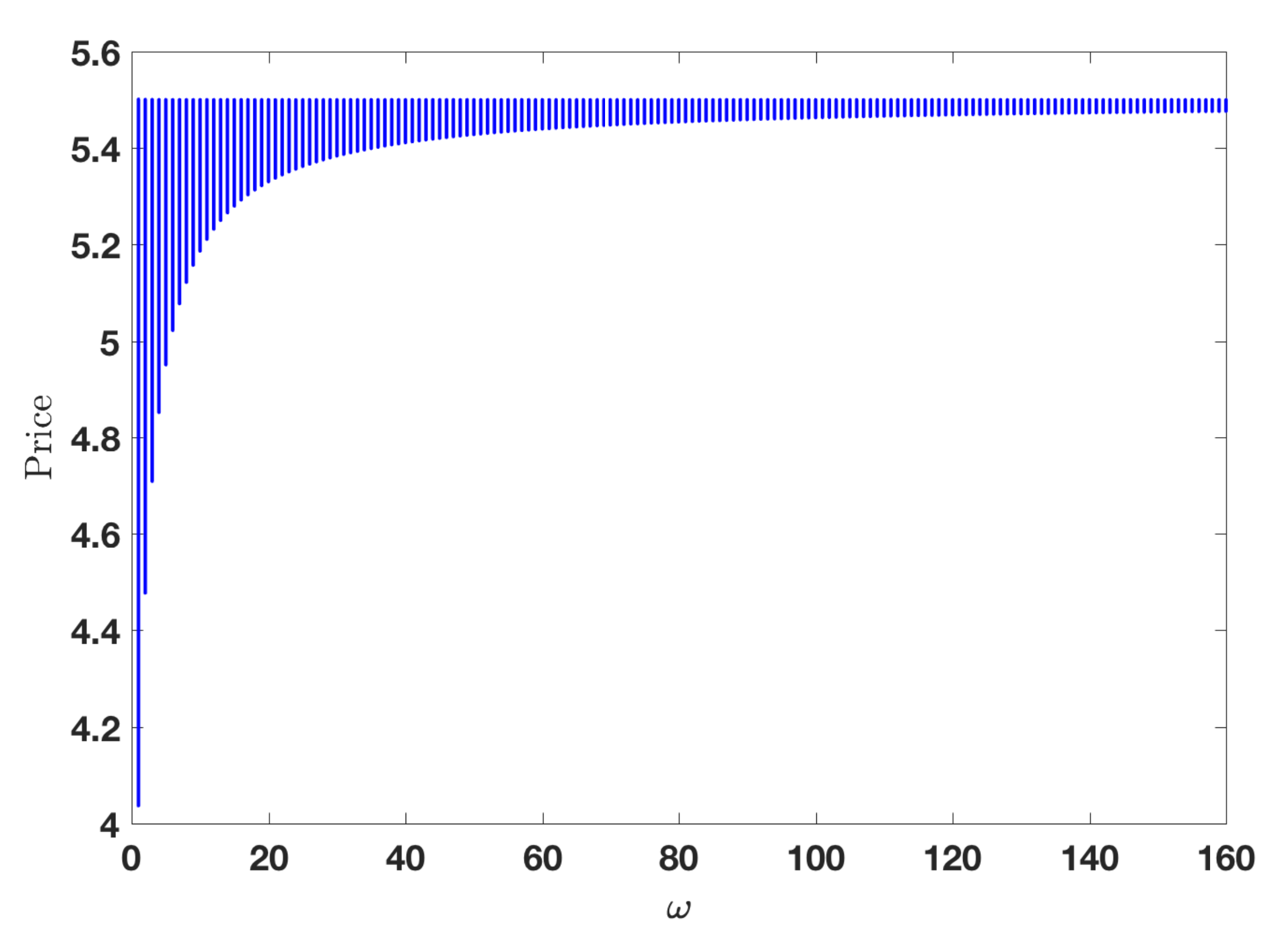}
\hfill
\includegraphics[scale=0.3]{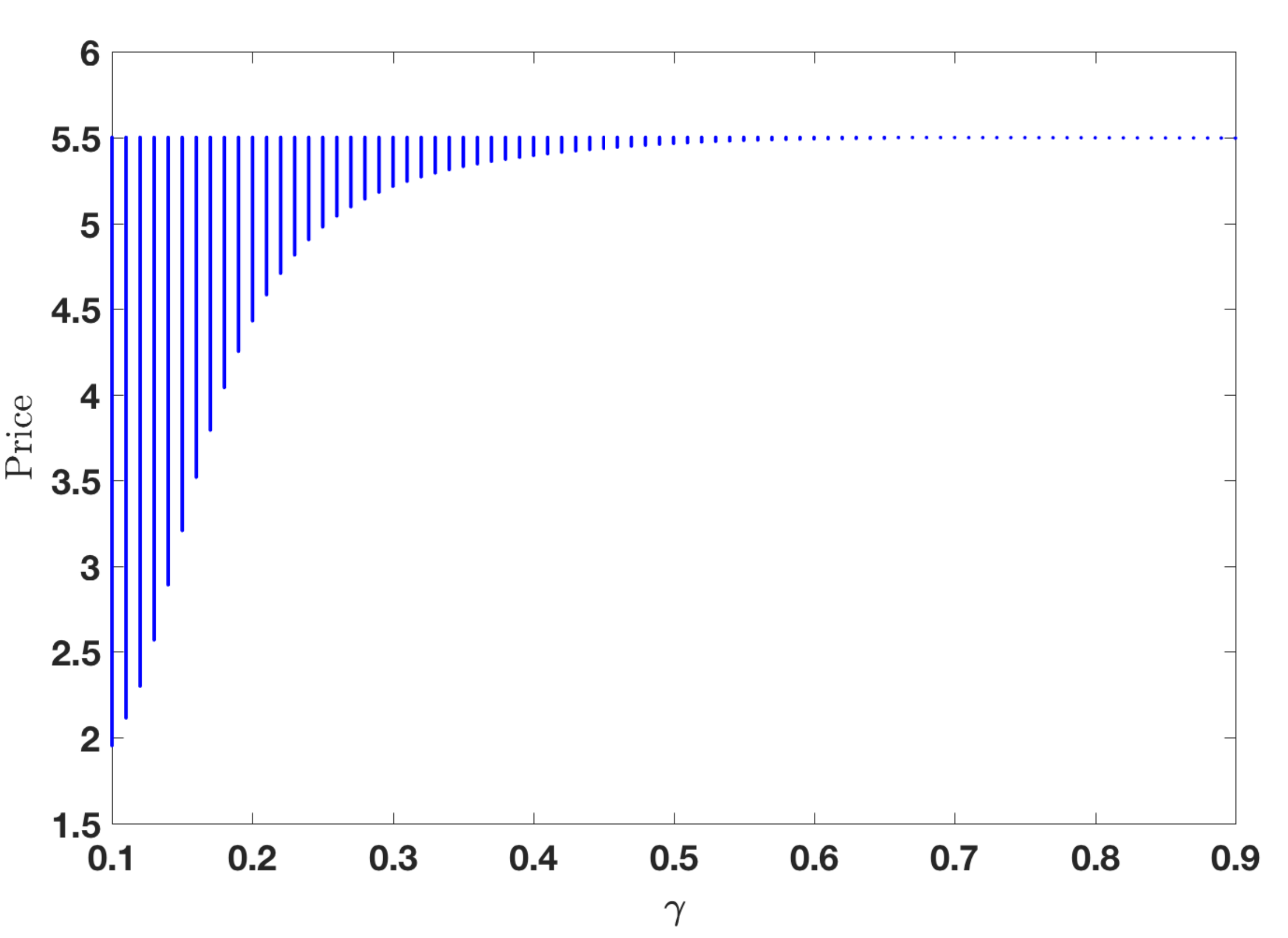}
\end{center}
\caption{Asymptotic stock price behavior for varying $\omega$ (left hand side) and $\gamma$ (right hand side). For further parameter settings we refer to table  \ref{Param} in appendix \ref{paramMom}.} \label{Asymp1}
\end{figure}

\begin{figure}[h!]
\begin{center}
\includegraphics[scale=0.3]{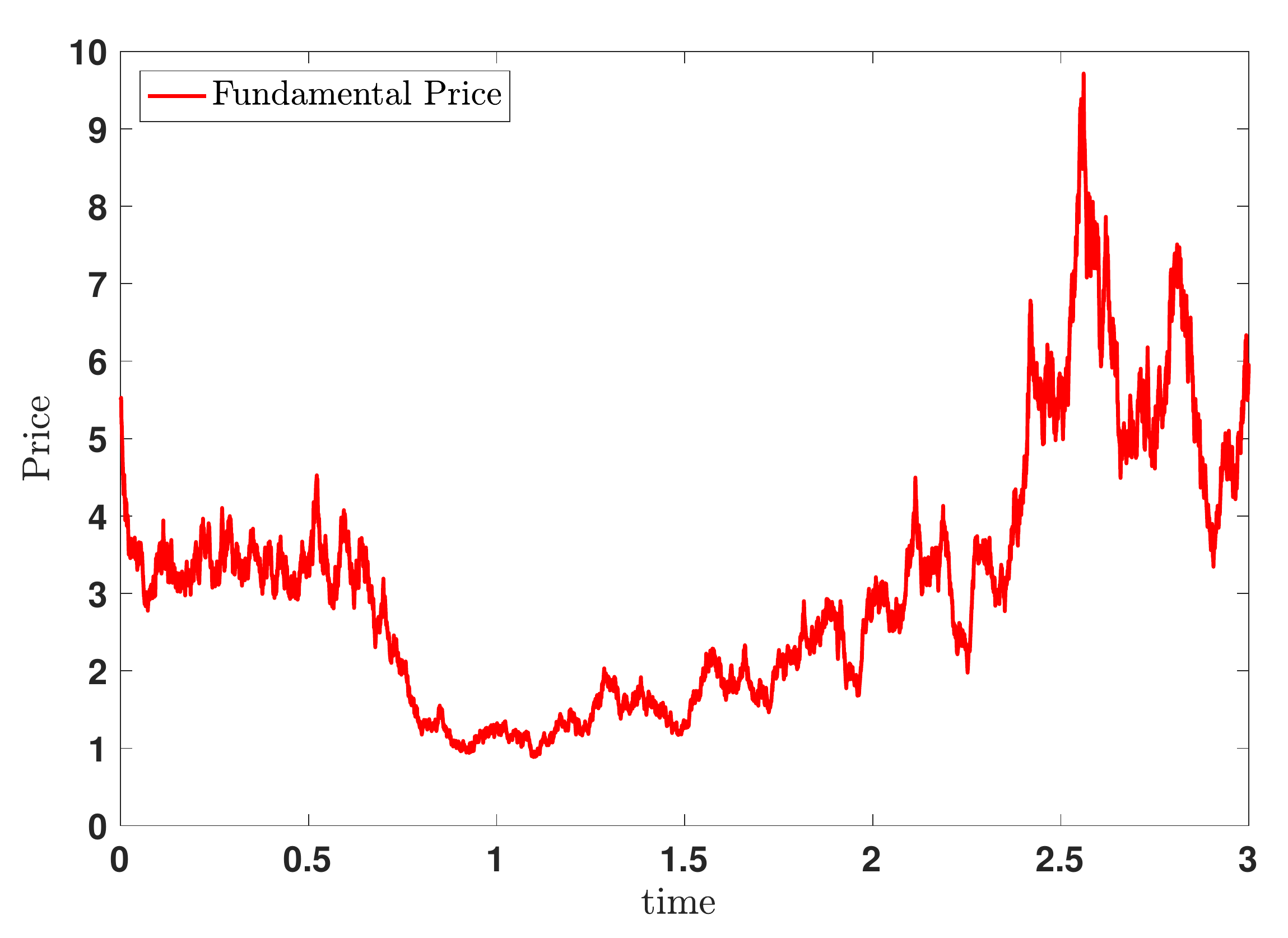}
\hfill
\includegraphics[scale=0.3]{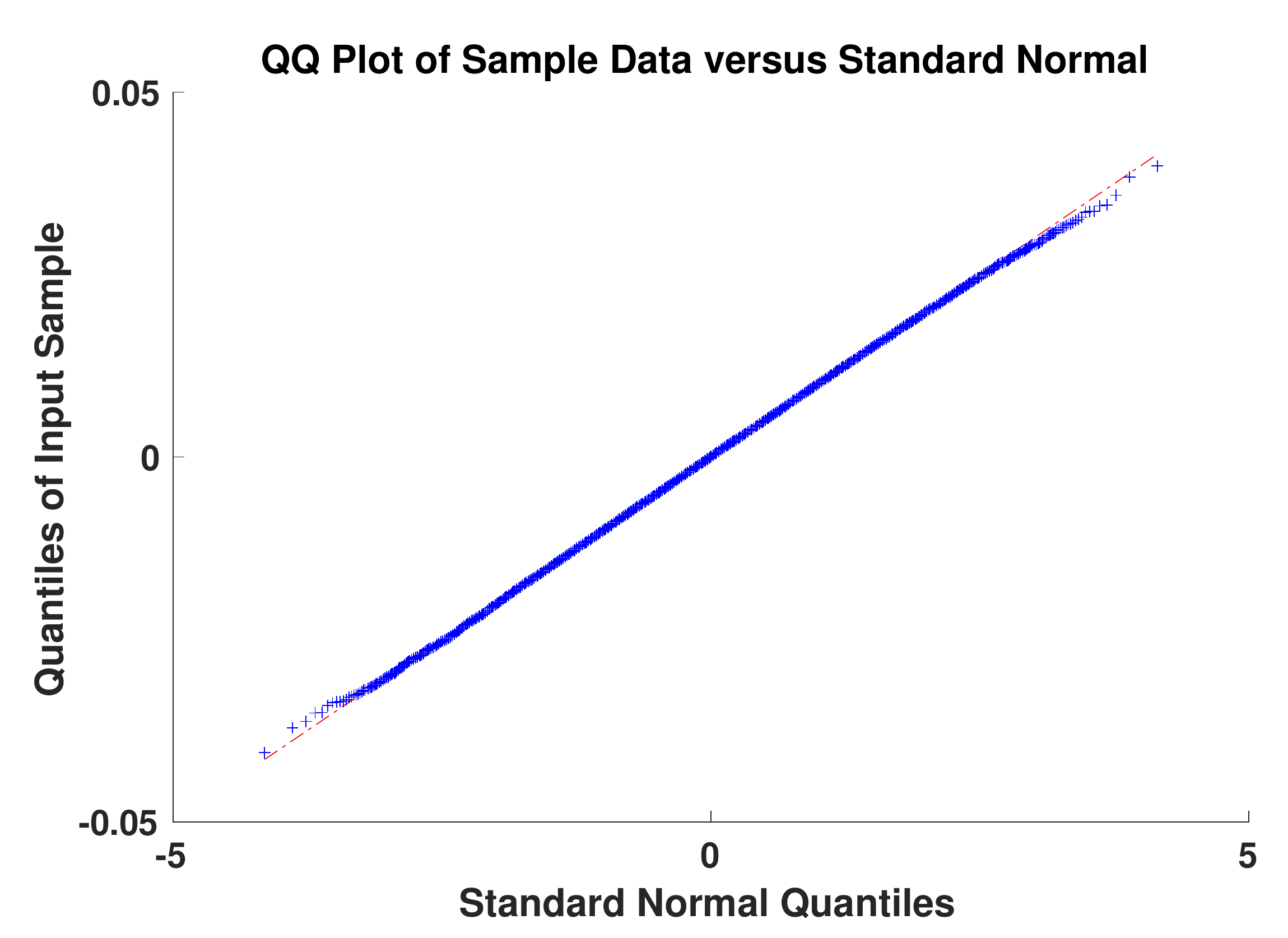}
\end{center}
\caption{Random fundamental price $s^f$ (left hand side) and the corresponding logarithmic return (right hand side) in a quantile-quantile plot fitted by a Gaussian distribution. Initial fundamental price is set to 
$s^f_0=5.5$ and the Matlab random seed \texttt{rng(767)} was chosen. }  \label{stochasticF}
\end{figure}

\begin{figure}[h!]
\begin{center}
\includegraphics[scale=0.3]{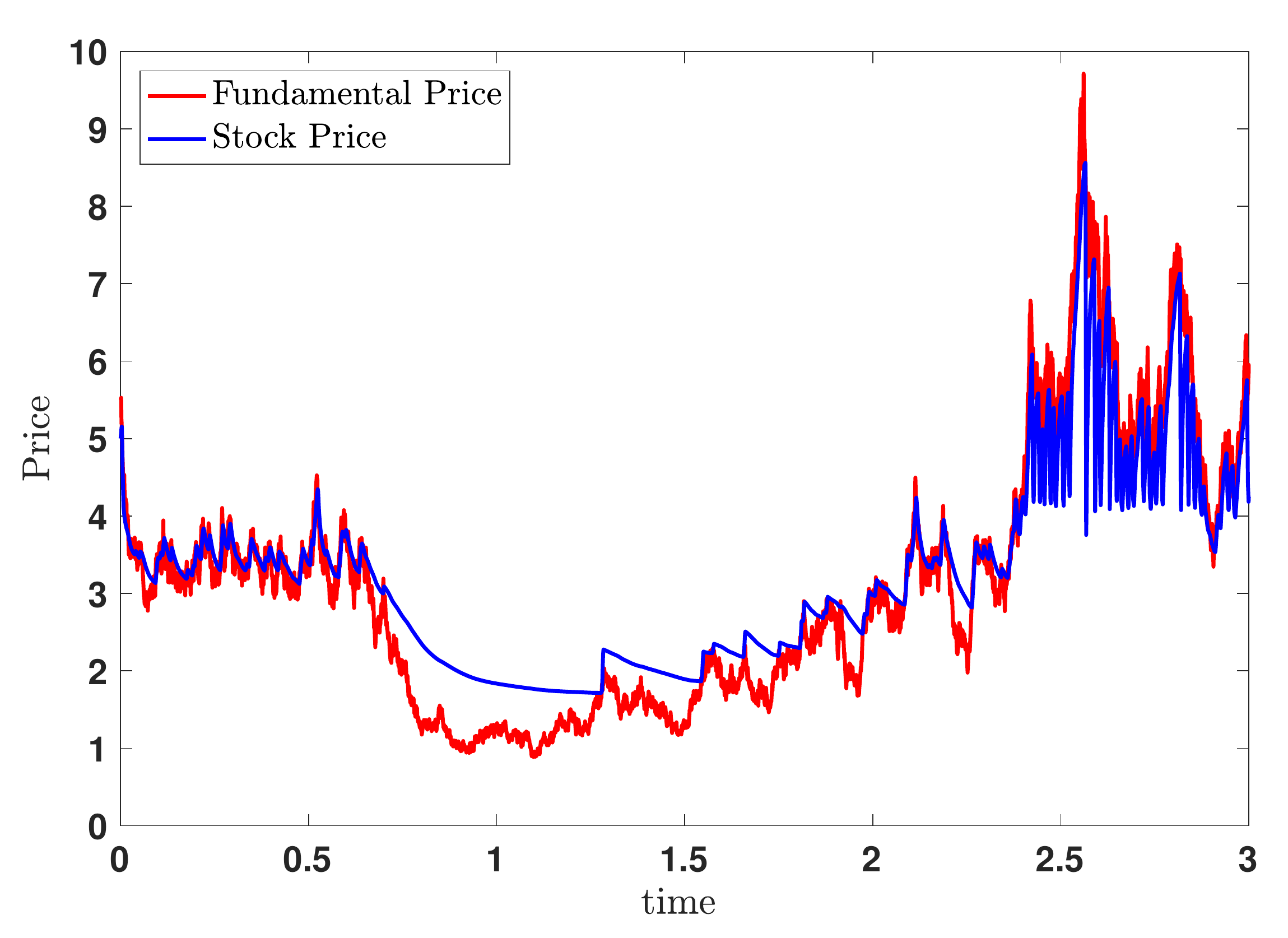}
\hfill
\includegraphics[scale=0.3]{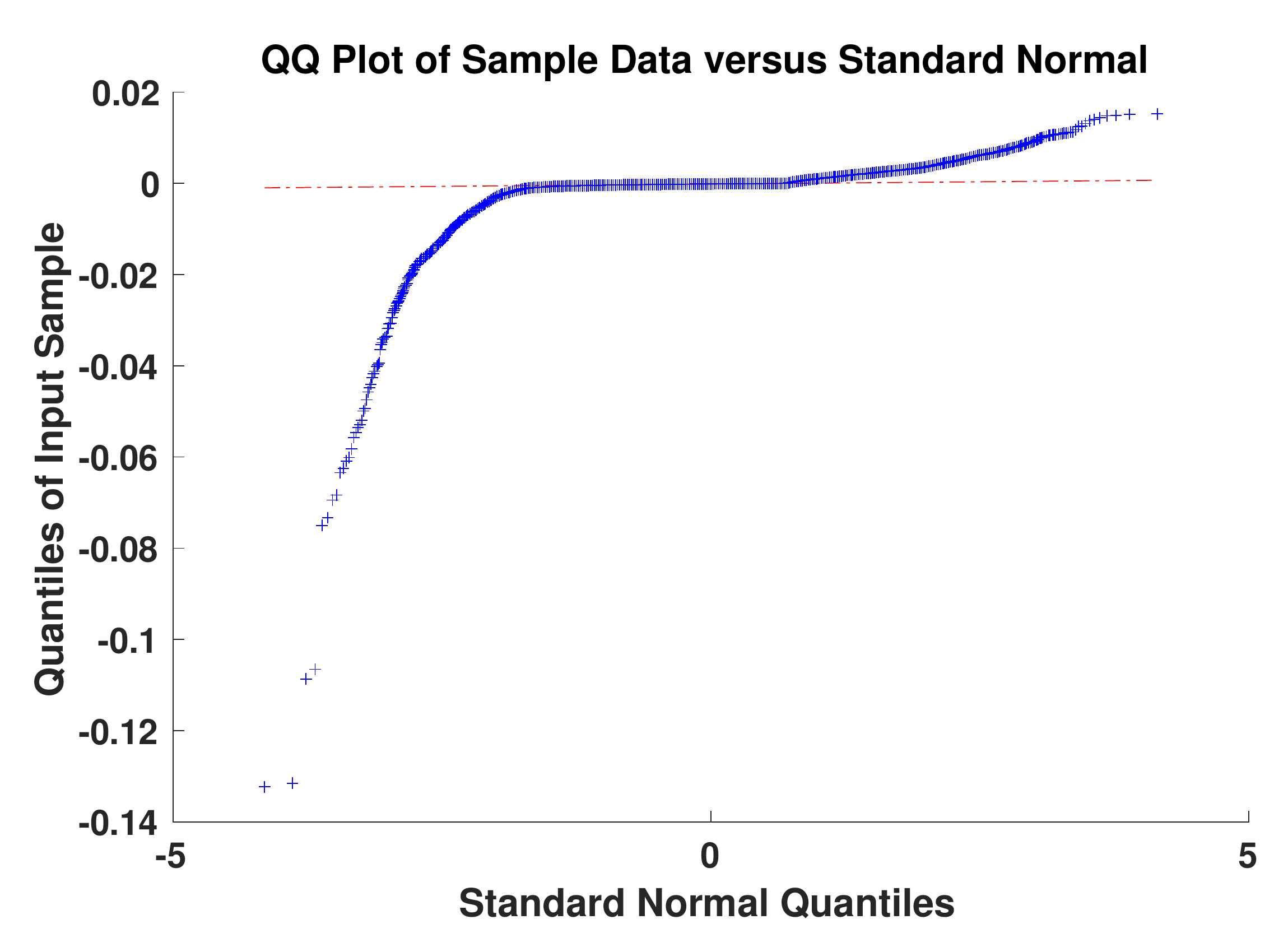}
\end{center}
\caption{The simulation was conducted with the parameters $\gamma=0.9, \omega=160,\ \beta=0.25$. Further parameters were chosen accordingly to table \ref{Param} in appendix \ref{paramMom}.}\label{RealStock}
\end{figure}

\begin{figure}[h!]
\begin{center}
\includegraphics[scale=0.3]{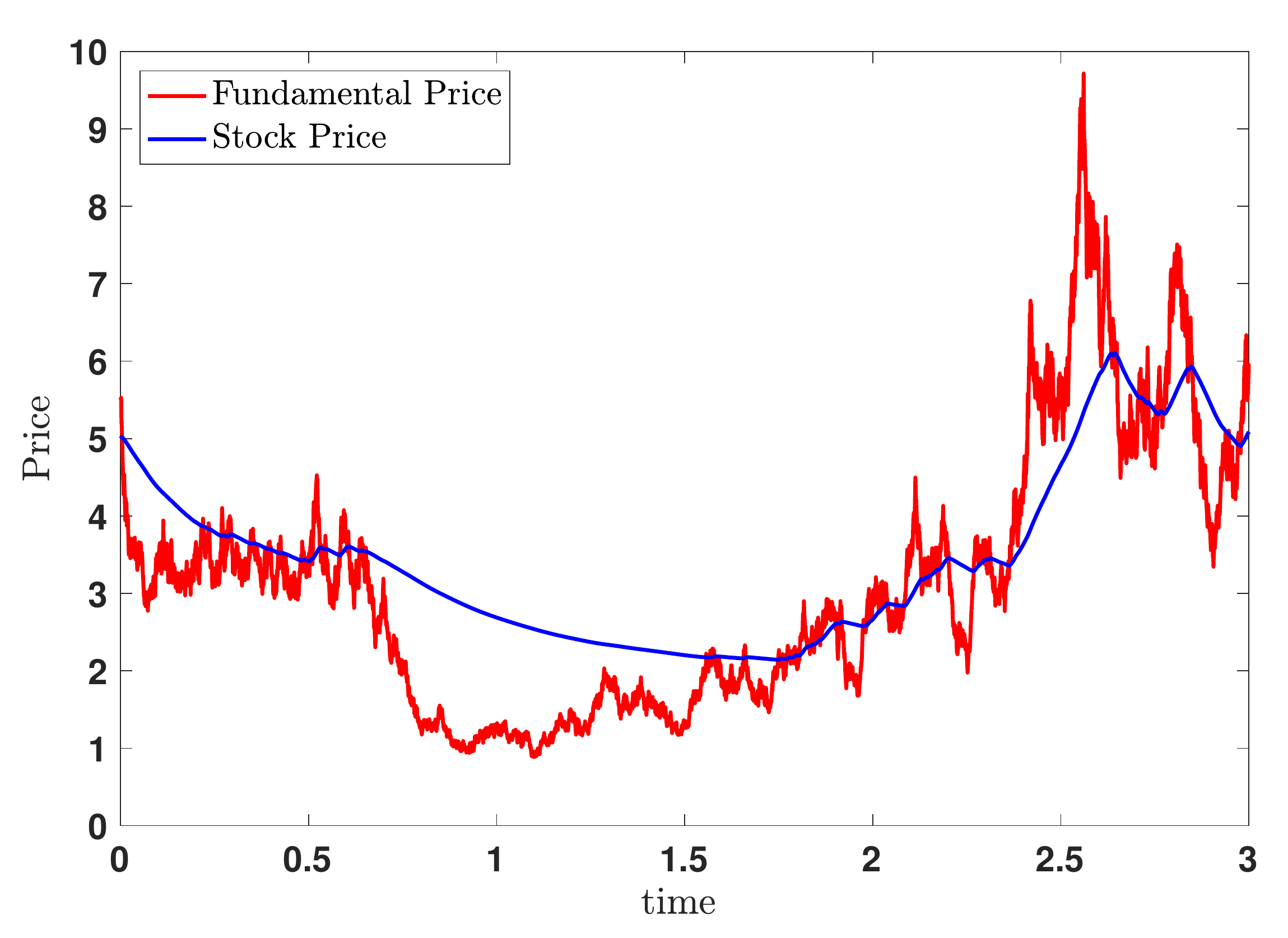}
\hfill
\includegraphics[scale=0.3]{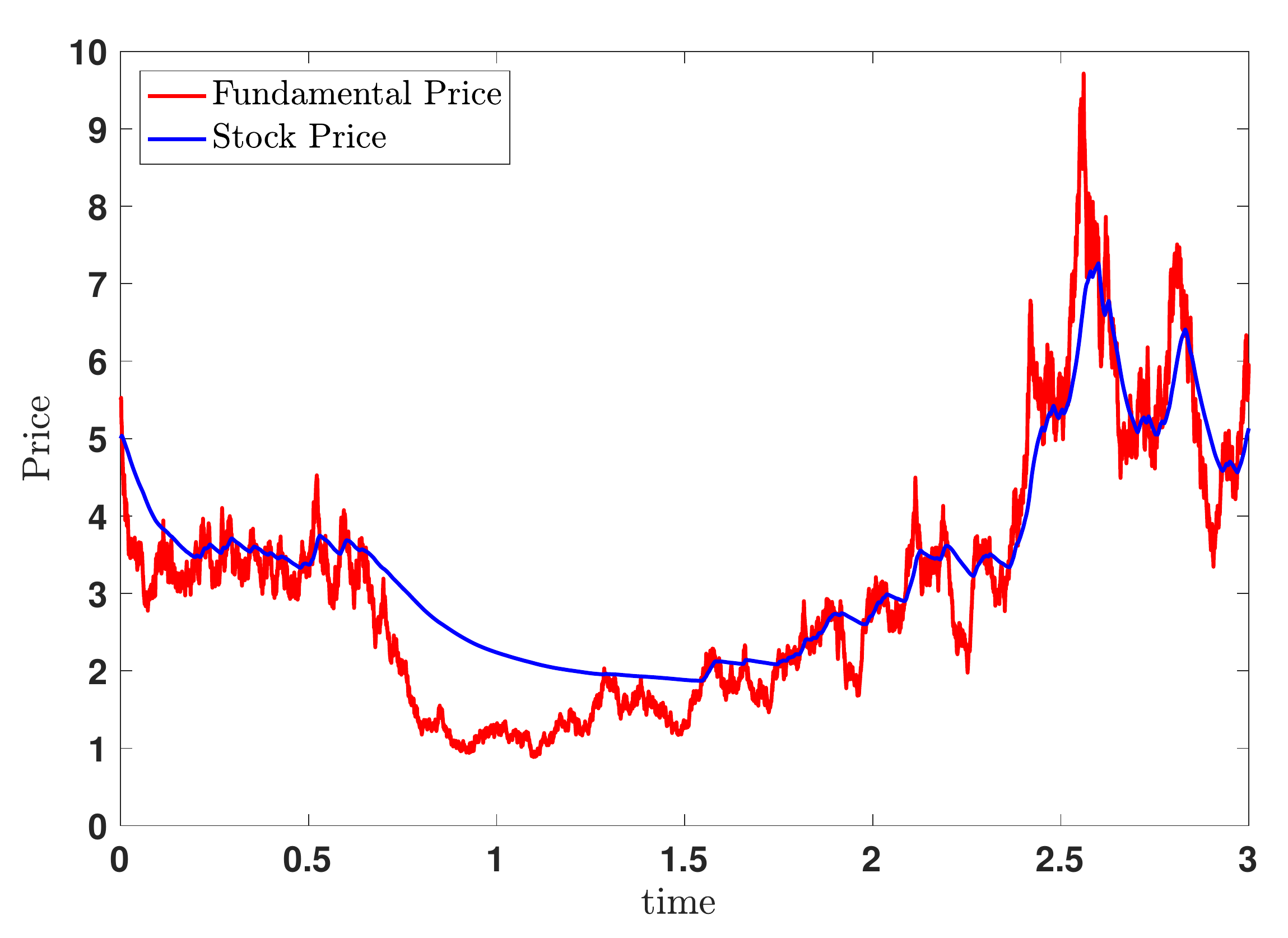}
\end{center}
\caption{ We have chosen the same parameter values as in the simulations in Figure \ref{RealStock} except the risk tolerance $\gamma$. In the figure on the left hand side we have chosen a risk tolerance of $\gamma=0.45$ and in the figure on the right hand side $\gamma=0.65$.}\label{DiffGamma}
\end{figure}

In order to quantify the findings of the previous Remark $\ref{remark}$ we analyze the asymptotic behavior of the stock price with respect to the parameters $ \omega, \gamma$. 
In fact we have simulated the model for $700,000$ time steps. In Figure \ref{Asymp1} we have plotted the parameter value against the stock price ranges of the last $200,000$
time steps. A dot corresponds to a converged stock price (steady state), whereas a line to oscillatory stock prices. Hence, Figure \ref{Asymp1} reveals that an increase in $\omega$ and $\gamma$
leads to a decrease of the wave amplitude. 
%\begin{figure}[h!]
%\begin{center}
%\end{center}
%\caption{Asymptotic stock price behavior for varying $\gamma$. For further parameter settings we refer to table  \ref{Param} in appendix \ref{paramMom}. } \label{Asymp2}
%\end{figure}
%
%
%
\paragraph{Random fundamental price}
Although the deterministic model can reproduce booms and crashes, the periodic behavior is very unrealistic. 
In order to obtain reasonable stock price data we introduce a random fundamental price. The fundamental price is defined as the solution 
of the SDE
$$
ds^f= s^f\ dW, \quad s^f(0)=s^f_0,
$$
which needs to be interpreted in the It\^o sense. For our numerical investigations we have simply added the SDE to the macroscopic portfolio model. 
In fact it would be also possible to add the previously introduced SDE to the microscopic model. Then one needs to repeat the MPC formalism in the stochastic setting, which is in general possible. \\ \\
The logarithmic return distribution of the stochastic process $s^f$ is well fitted by a Gaussian distribution (see Figure \ref{stochasticF}).
The Figure \ref{RealStock} reveals that we obtain realistic stock price data with a random fundamental price. The quantile-quantile plot in Figure \ref{RealStock} clearly illustrates the existence of fat tails in the logarithmic stock price return distribution. In addition, we may note that the stock price usually follows the fundamental price, but it is also possible to obtain market regimes where the stock price is more volatile than the fundamental price. Furthermore, the Figure \ref{DiffGamma} shows that a decreasing risk tolerance leads to slower chasing of the fundamental price by the stock price and a less volatile price behavior.

\begin{remark}
Different choices of the weighting function $W$ and value function $U_{\gamma}$  have led to similar oscillatory behavior.
Certainly the influence and impact of parameters may change for varied functions $W$ and $U_{\gamma}$.
\end{remark}

\section{Concluding Discussion}

In this work we have established a macroscopic portfolio model, which can be derived from microscopic agent dynamics. On the microscopic level we have introduced an approximation framework of the optimal control problem in order to give a precise definition of rational and bounded rational agents. The model can be regarded as a model with fully myopic agents. The qualitative discussion has led to the conjecture that an interplay of chartist and fundamental trading behavior is essential in order to obtain oscillatory, cyclic price behavior. Our simulations confirm the existence of cyclic price behavior around the fundamental value. Thus, the model offers the mean reversion characteristic. This model behavior is similar to the stock price behavior in the models \cite{brock1998heterogeneous, chiarella1992dynamics, lux1995herd}. Interestingly, the trust coefficient $\beta$ determines the location of the oscillations with respect to the fundamental price. This indicates that the agent's attitude towards the currently best performing trading strategy play a major role in the price formation. Such a trust parameter has not been introduced by any earlier model and it may be worthwhile to study the impact of that parameter in more detail in the future. In addition, the parameter studies reveal that the risk tolerance $\gamma$ and the reaction strength $\omega$ of fundamentalists heavily influence the wave speed and amplitude. In our case a higher risk tolerance of investors leads to less volatile prices and less pronounced booms and crashes. The reason is that an increasing risk tolerance
leads to a larger impact of the fundamental trading strategy. Similar model behavior has been reported in \cite{chiarella2002heterogeneous}. It has been pointed out by Odean \cite{odean1998volume} that a small risk tolerance of investors causes overconfidence and this can cause higher volatility. The reason is that trader underweight rational information which are in our model given by the fundamental trading strategy. Furthermore, the simulations conducted with random fundamental prices have led to realistic price movements. Especially, we could obtain fat tails in logarithmic asset returns. \clearpage
Apart from the dynamic behavior of our macroeconomic portfolio model which coincides with simulations of other asset pricing models, we want to stress two points. 
\begin{itemize}
\item The resulting marcoeconomic portfolio model has been derived from microscopic agent dynamics and the myopic trading rules can be seen as a simple approximation of a very elaborated optimal control problem in the context of differential games. 
\item Thanks to model predictive control, we were able to give a precise mathematical definition of the degree of rationality of the financial agent. This has been done for a rather general portfolio model and one can even expect to apply this methodology to other optimal control problems in finance. 
\end{itemize}
Nevertheless, we have to admit that we consider a simple portfolio model. Various extensions and generalizations are possible. One may add uncertainty in the microscopic portfolio model or study the impact of increasing rationality of the microscopic agents on the model dynamics. 
In addition, it seems interesting to study the impact of different cost functions on the stock price behavior. Especially a rigorous quantification of the oscillatory stock price behavior is of major interest.\\ \\
This work shows that heuristic trading strategies of investors can be interpreted as approximations of investments by a rational financial agent. Although a fully bounded rational agent seems to be quite unrealistic, the mathematical connection to a perfect rational agent may help to discover new and more appropriate models of financial agents.

\section*{Acknowledgement}
Torsten Trimborn was supported by the Hans-B\"ockler-Stiftung. 
\clearpage

\appendix
\section*{Appendix}
\renewcommand{\thesubsection}{\Alph{subsection}}
\subsection{}\label{compMom}

\paragraph{Qualitative analysis}
The proof of proposition \ref{prop2} reads:
\begin{proof}
We show local Lipschitz continuity. Then existence and uniqueness directly follows by the Picard-Lindel\"of theorem. 
We can rewrite the stock price equation into an explicit ODE system:
\begin{align}\label{expStock}
\dot{S} = 
\begin{cases}
\kappa\ (\chi\omega s^f+(1-\chi)\ D-(r+\chi \omega)\ S)\ \frac{Y}{1+(1-\chi)\kappa\ Y},\quad K(S)>0,\\
\kappa\ (\chi\omega s^f+(1-\chi)\ D-(r+\chi \omega)\ S)\ \frac{X}{1+(1-\chi)\kappa\ X},\quad K(S)<0.
\end{cases}
\end{align}
Thus, we may denote the right hand side of the explicit ODE system by $F(z),\ z:=(X,Y,S)^T \in[0,\infty)\times [0,\infty)\times (0,\infty)$. Notice that the excess demand $ED$ does no longer depend on $\dot{S}$, since we can insert the right hand side of the stock price equation \eqref{expStock}. Local Lipschitz continuity is obvious except for the potential singularity in  $S^*=\frac{\chi\ \omega\ s^f+(1-\chi)\ D}{r+\chi\ \omega}$, since $K(S^*)=0$ holds. Thus, we show 
Lipschitz continuity on $U=U_X\times U_Y\times U_S, \  U_X:= [X_0-\epsilon, X_0+\epsilon],\ U_Y:= [Y_0-\epsilon, Y_0+\epsilon],\ U_S:= [S^*-\epsilon, S^*+\epsilon],\ \epsilon>0 $ with $z_0:= (X_0, Y_0, S^*)$, where $X_0,Y_0\in [0,\infty)$ are arbitrary but fixed.  First we discuss the excess demand $ED$:
\begin{align*}
|ED(X,Y,S)| \leq & \kappa\ (r+\chi\ \omega)\  \max\left\{   \max\limits_{X \in U_X} |X|,  \max\limits_{Y \in U_Y} |Y|  \right\}\ \max\limits_{S\in U_S} \left| \frac{1}{S}\right|  | S^*-S | \\
 &\left( 1+ \kappa\  (1-\chi)\ \max \left\{ \max\limits_{X \in U_X} \left| \frac{X}{1+(1-\chi)\kappa\ X} \right|, \max\limits_{Y \in U_Y} \left| \frac{Y}{1+(1-\chi)\kappa\ Y} \right|  \right\}   \right)\\
 &\leq  C_1\ |S^*-S|
\end{align*}
As next step we discuss each component of $F=(F_1,F_2,F_3)^T$ separately. For the stock price evolution we obtain:
\begin{small}
\begin{align*}
|F_3(Z)-F_3(Z_0)| & \leq \kappa\ (r+\chi \omega)\ \max \left\{ \max\limits_{X \in U_X} \left| \frac{X}{1+(1-\chi)\kappa\ X} \right|, \max\limits_{Y \in U_Y} \left| \frac{Y}{1+(1-\chi)\kappa\ Y} \right|  \right\} \ |S^*-S|\\
&\leq C_2\  |S-S^*|
\end{align*}
\end{small}
For the portfolio dynamics we get:
\begin{align*}
|F_1(Z)-F_1(Z_0)| \leq & D\ \max\limits_{S\in U_S} \left| \frac{1}{S} \right|\ |X-X_0|+ (1+\kappa\ \max\limits_{X\in U_X} |X|)\ C_1\ |S^*-S|\\
  \leq & C_3\ |X-X_0|+ C_4\ |S-S^*|,\\
|F_2(Z)-F_2(Z_0)| \leq & r\ |Y-Y_0|+ \ C_1\ |S^*-S|.
\end{align*}
Hence, we conclude that 
$$
||F(z)-F(z_0)||\leq L\ ||z-z_0 ||,\quad z,z_0 \in U,
$$
holds on $U$ with Lipschitz constant $L:= C (C_1+C_2+C_3+ C_4 +r)$, where the additionally constant $C$  is due to the equivalence of norms. 
\end{proof}

The proof of proposition \ref{prop3} is given by:
\begin{proof}
We set $r=D=0$ and derive the explicit ODE system.
Thus, for a continuous differentiable Lyapunov functional we can compute the Lie derivative. 
We define the Lyapunov functional as follows: $\psi: \R^3 \to \R,\ (x,y,S)^T\mapsto -(s^f-S)\ (x+y-(s^f-s))$. We immediately obtain
$$
\frac{d}{dt}\psi((X(t),Y(t),S(t)))\leq 0,\quad \text{with}\ \frac{d}{dt}\psi((X(t),Y(t),S(t)))\Big|_{S=S_{\infty}}=0,
$$
and can conclude the asymptotic stability of $S_{\infty}$.
\end{proof}

\begin{proposition}
In special cases, we can compute solutions of the stock price equation. We assume constant weights $\chi$ and assume that the utility function is described by the identity.  
\begin{itemize}
\item Fundamentalists alone ($\chi=1$): The stock price equation reads
$$
\dot{S} =\begin{cases}
 \kappa\ (\omega\ s^f- (\omega+r)\ S)\ Y,\quad  \frac{\omega s^f}{\omega+r}>S,\\
 \kappa\ (\omega\ s^f-(\omega+r)\ S)\  X,\quad  \frac{\omega s^f}{\omega+r}<S . 
\end{cases}
$$
This equation seems reasonable, so the investor shifts his capital into stocks if he expects a positive stock return, and vice versa. The solution is given by
$$
S(t)= 
\begin{cases}
(1-\exp\{- \kappa\ (\omega+r)\ \int\limits_0^t Y(\tau)\ d\tau )\})\ \frac{\omega\ s^f}{\omega+r} +S(0) \ \exp\{- \kappa \ (\omega+r)\ \int\limits_0^tY(\tau)\ d\tau\},\\
\quad \text{for}\  \frac{s^f}{\omega+r}>S,\\
(1-\exp\{- \kappa\ (\omega+r)\ \int\limits_0^t X(\tau)\ d\tau )\})\ \frac{\omega\ s^f}{\omega +r} +S(0) \ \exp\{- \kappa\  (\omega+r)\ \int\limits_0^tX(\tau)\ d\tau\},\\
\quad\text{for}\ \frac{s^f}{\omega+r}<S.
\end{cases}
$$
Hence, the price is driven exponentially fast to the steady state $S_{\infty}=\frac{\omega\ s^f}{\omega+r}$.
\item Chartists alone ($\chi = 0$):  We get
\begin{align*}
\dot{S } = \begin{cases}
\frac{\kappa\ D\ Y}{1-\kappa\ Y}- \frac{r\ \kappa Y}{1-\kappa Y}\ S,\quad  \text{for}\ D>S,\\
\frac{\kappa\ D\ X}{1-\kappa\ X}- \frac{r\ \kappa X}{1-\kappa X}\ S,\quad  \text{for}\ D<S.
\end{cases}
\end{align*}
The solution is given by 
\begin{footnotesize}
\begin{align*}
S(t)= \begin{cases}
\left( 1-\exp\left\{  -r \ \kappa \ \int\limits_0^t \frac{Y(\tau)}{1-\kappa Y(\tau)}\ d\tau  \right\}\right)\ \frac{D}{r}+S(0)\ \exp\left\{  -r \ \kappa \ \int\limits_0^t \frac{Y(\tau)}{1-\kappa Y(\tau)}\ d\tau  \right\},\\ 
\quad \text{for}\ \kappa\ D\ Y +D\ (1-\kappa\ Y)>S,\\
\left(1+  \exp\left\{  -r \ \kappa\ \int\limits_0^t \frac{X(\tau)}{1-\kappa\ X(\tau)}\ d\tau  \right\}\right)\ \frac{D}{r}+S(0)\ \exp\left\{  -r \ \kappa\ \int\limits_0^t \frac{X(\tau)}{1-\kappa\ X(\tau)}\ d\tau  \right\},\\
\quad \text{for}\  \kappa\ D\ X+D\ (1-\kappa\ Y)<S.
\end{cases}
\end{align*}
\end{footnotesize}
\item Chartists and fundamentalists with a constant weight $\chi\in (0,1)$: 
The corresponding stock price equation reads
\begin{align*}
\dot{S} = 
\begin{cases}
\kappa\ (\chi\omega s^f+(1-\chi)\ D-(r+\chi \omega)\ S)\ \frac{Y}{1+(1-\chi)\kappa\ Y},\quad K(S)>0,\\
\kappa\ (\chi\omega s^f+(1-\chi)\ D-(r+\chi \omega)\ S)\ \frac{X}{1+(1-\chi)\kappa\ X},\quad K(S)<0.
\end{cases}
\end{align*}
The solution is given by
\begin{footnotesize}
$$
S(t)= 
\begin{cases}
(1-\exp\{- \kappa\ (\chi \omega+r)\ \int\limits_0^t \frac{Y(\tau)}{1+\kappa (1-\chi) Y(\tau)}\ d\tau )\})\ \frac{\chi \omega\ s^f+(1-\chi)\ D}{\chi \omega+r} \\
\quad +S(0) \ \exp\{- \kappa \ (\chi \omega+r)\ \int\limits_0^t\frac{Y(\tau)}{1+\kappa (1-\chi) Y(\tau)}\ d\tau\},\quad \text{for}\ K(S)>0,\\
(1-\exp\{- \kappa\ (\chi \omega+r)\ \int\limits_0^t \frac{X(\tau)}{1+\kappa (1-\chi) X(\tau)}\ d\tau )\})\ \frac{\chi \omega\ s^f+(1-\chi)\ D}{\chi \omega +r} \\
\quad +S(0) \ \exp\{- \kappa\  (\chi \omega+r)\ \int\limits_0^t\frac{X(\tau)}{1+\kappa (1-\chi) X(\tau)}\ d\tau\},\quad \text{for}\ K(S)<0.
\end{cases}
$$
\end{footnotesize}
\end{itemize}
\end{proposition}

\begin{proposition}
For the wealth evolution, we consider the stock and bond portfolio separately.
\begin{itemize}
\item In the stock portfolio, the wealth evolution is given by
$$
\dot{X} = 
\begin{cases}
(\kappa\ K(S)\ Y+ \frac{D}{S})\  X + K(S)\ Y,&\quad \text{for}\ K(S)>0,\\
(\kappa\ K(S)\ X +\frac{D}{S})\  X+ K(S)\ X,&\quad\text{for}\ K(S)<0.
\end{cases}
$$
The solution is given by 
$$
X(t) = 
\begin{cases}
X(0)\exp\left\{ \int\limits_0^t\kappa\ K(S) Y\ + \frac{D}{S}\ d\tau    \right\}+\left(1 - \exp\left\{-\int\limits_0^t\kappa \ K(S) Y\ d\tau  \right\}\right)\frac{1}{\kappa},\\
\quad \text{for}\ K(S)>0,\\
\frac{X(0)\ \exp\left\{ \int\limits_0^t K(s)+\frac{D}{S}\ d\tau  \right\} }{ 1+\kappa \int\limits_0^t K(S)\ \exp\left\{ \int\limits_0^{\zeta} K(S)+\frac{D}{S}\ d\zeta\right\}\ d\tau },\quad \text{for}\ K(S)<0.
\end{cases}
$$
\item The bond portfolio is given by 
$$
\dot{Y} = 
\begin{cases}
r \ Y - K(S)\ Y,\quad &\text{for}\ K(S)>0,\\
r\  Y - K(S)\ X,\quad &\text{for}\ K(S)<0, 
\end{cases}
$$
with the solution
$$
Y(t) = 
\begin{cases}
Y(0)\ \exp\left\{ \int\limits_0^t (r-K(S))\ d\tau  \right\},&\quad\text{for}\ K(S)>0,\\
\exp\left\{ r\ t   \right\}\left( Y(0)-
\int\limits_0^t K(S)\ X \exp\left\{-r \ \tau \right\}\ d\tau\right),&\quad \text{for}\ K(S)<0.
\end{cases}
$$
\end{itemize}
\end{proposition}

\newpage

\subsection{}\label{paramMom}

\paragraph{Parameters of simulation}
If not indicated differently the parameters are set to:

\begin{figure}[h!]
\begin{center}
\begin{tabular}{|c|c||c|c|}
\hline
$\Delta t$ & $0.0001$ & $\kappa$ & $0.1$ \\
\hline
$D$ &  $0.01$ & $\nu$ &$ 5$ \\
\hline
$r$ &  $0.01$ & $S_0$ & $ 5$ \\
\hline
$\alpha $ &  $0.5$ & $Y_0$ & $ 20$ \\
\hline
$\omega$ &  $20$ & $X_0$&  $20$ \\
\hline
$ s^f$ &  $5.5$ &$T_{end}$ & $3$   \\
\hline
$ \gamma$ &  $0.35$ & &     \\
\hline
\hline
\end{tabular} 
\caption{Standard parameter setting.}\label{Param}
\end{center}
\end{figure}

%-- LITERATUR ----------------------------------------------------------%
	\clearpage
	\bibliography{literaturmean.bib}
		\bibliographystyle{abbrv}

\end{document}